\documentclass[preprint,12pt,authoryear]{elsarticle}

\usepackage{amssymb}
\usepackage{amsmath}
\usepackage{bm}
\usepackage{mathtools}  
\usepackage{float}      
\usepackage{afterpage}  

\usepackage{booktabs}  
\usepackage{caption}   
\usepackage{subcaption}

\usepackage{natbib}
\usepackage{hyperref}
\hypersetup{
    colorlinks=true,
    citecolor=blue,
    linkcolor=blue,
    filecolor=magenta,      
    urlcolor=cyan,
}

\journal{arXiv}

\begin{document}

\begin{frontmatter}



\title{Physics-Informed Machine Learning for Grade Prediction in Froth Flotation}



\author[label1]{Mahdi Nasiri}
\author[label1]{Sahel Iqbal}
\author[label1]{Simo Särkkä}
\affiliation[label1]{organization={Aalto University},
            addressline={Rakentajanaukio 2 C},
            city={Espoo},
            postcode={02150},
            state={Uusima},
            country={Finland}}



\begin{abstract}
In this paper, physics-informed neural network models are developed to predict the concentrate gold grade in froth flotation cells. Accurate prediction of concentrate grades is important for the automatic control and optimization of mineral processing. Both first-principles and data-driven machine learning methods have been used to model the flotation process. The complexity of models based on first-principles restricts their direct use, while purely data-driven models often fail in dynamic industrial environments, leading to poor generalization. To address these limitations, this study integrates classical mathematical models of froth flotation processes with conventional deep learning methods to construct physics-informed neural networks. These models demonstrated superior generalization and predictive performance compared to purely data-driven models, on simulated data from two flotation cells, in terms of mean squared error and mean relative error.

\end{abstract}







\begin{keyword}



machine learning \sep physics-informed neural networks \sep froth flotation \sep predictive modeling
\end{keyword}

\end{frontmatter}



\section{Introduction}
\label{app:intro}

The primary objective of mineral processing is to liberate valuable minerals from the gangue~\citep{Wills2015}. This is achieved by first reducing the ore size, and then by physically separating and concentrating the desirable minerals through various methods, among which froth flotation is the most important and complex~\citep{AZHIN2021107028, Radmehrmineral}. In froth flotation, the concentrate grade is a key indicator of production success~\citep{Wang2015}. On-line prediction of grade is essential for accurate control and optimization of industrial processes~\citep{Wang2024}. While direct measurement and estimation of this variable often require instruments that are costly to acquire and maintain, soft sensor modeling, as the more cost-effective and faster method, may be employed to predict concentrate grade based on variables that can be measured on-line~\citep{MORAR2012, JAHEDSARAVANI2014, Ren2015}. The predicted variable can then be utilized as input for a feedback control system that adjusts the process variables to ensure optimal flotation performance~\citep{JAHEDSARAVANI201690, Wang2024}.

The modeling of froth flotation is challenging due to its complexity and the interaction of numerous variables~\citep{Quintanilla2021a}. Models range from empirical, such as artificial neural networks (ANN), to phenomenological, including probabilistic, kinetic, and population-balance models~\citep{Jovanovic2015}. However, no model can fully capture the entire dynamics of the process, often leading to significant control challenges~\citep{Quintanilla2021b, SCHWARZ2019106033}.

Data-driven models, such as those utilized in~\citet{GOMEZFLORES2022, Nakhaei2013} are efficient at identifying hidden patterns and modeling nonlinear relationships but encounter notable challenges. These models are highly dependent on specific datasets, struggle with adapting to dynamic and noisy industrial environments, and require extensive data that are seldom met in real-world applications~\citep{karniadakis2021physics, willard2022integrating}. Moreover, they typically lack physical interpretability, which limits their ability to extrapolate and generalize beyond the data used for training~\citep{willard2022integrating, rajulapati2022integration}. On the other hand, physics-based models, although robust and broadly accepted, tend to simplify reality due to the complexities of real-world systems and our incomplete knowledge of the processes, potentially leading to biases~\citep{Jovanovic2015, willard2022integrating}. Additionally, the complexity of mathematical models used in mineral processing, particularly those applied to froth flotation~\citep{Wang2018CFDFlotation, Prakash2018FlotationTechnique, Dinariev2018DensityFunctional, Gharai2016ModelingFlotation, Wang2015EntrainmentReview, Jovanovic2015AdvancedControlFlotation, AlvesDosSantos2014ModellingFlotation}, constrains their direct application in control strategies, as they typically demand solutions that are both simple and robust enough to function effectively in real-time environments~\citep{Quintanilla2021b}.

One emerging approach to addressing these challenges is physics-informed machine learning, which integrates physical laws into machine learning algorithms. This approach is widely applied in various fields, including fluid dynamics~\citep{Cheng2021DeepLM, Haghighat2021PhysicsinformedNN, Mahmoud2022nnpinns}, kinetic processes~\citep{Weng2022MultiscalePINNs, Batuwatta2022moistureconcentrate}, and optimal control~\citep{Antonelo2021PhysicsInformedNN, Asrav2023PhysicsInformedRNN}. Combining domain knowledge with data can improve the performance of machine learning models by enabling them to utilize data for making precise predictions while also producing solutions that are physically interpretable, even in cases where the data may be noisy, sparse, or involve high-dimensional spaces~\citep{willard2022integrating, raissi2019physics}. These models have shown improvements in interpretability, extrapolation capabilities, and efficiency in computation, while offering additional advantages such as dimensionality reduction and superior generalization capabilities~\citep{rajulapati2022integration, hao2023physicsinformed}. Such methodologies could lead to more accurate control and decision-making in plant operations such as froth flotation, potentially increasing the adoption of deep learning and enhancing its applicability in dynamic environments~\citep{Azhari2023}.


The contribution of this work is the development of physics-informed neural networks to effectively model the dynamics of the froth flotation process. Specifically, this study leverages data from two froth flotation cells to predict concentrate gold grades by integrating classical mathematical models of the flotation process into conventional deep learning models. The datasets are generated from a digital twin model that continuously integrates real-process data with physics-based models calibrated using historical data. The data are first preprocessed to prepare them for model training, and three mathematical models, formulated as ordinary differential equations, are employed to develop physics-informed neural network (PINN) models. Finally, these models are compared against their purely data-driven counterparts, as well as traditional machine learning models such as linear regression (LR), random forest (RF), and decision tree (DT). Performance is assessed using metrics such as mean squared error (MSE) and mean relative error (MRE) to evaluate their effectiveness in managing sparse and noisy industrial data. The results demonstrate the superior generalization and predictive performance of the physics-informed neural network models over the purely data-driven models with respect to both mean squared error and mean relative error.
%
%
%
%

\section{Physics-Informed Neural Networks}
\label{app:PINN}
This section discusses the basic concepts of physics-informed neural networks, focusing on solving inverse problems.

Physics-informed neural networks (PINNs) are designed to model physical systems by leveraging available data along with known or partially known physical laws~\citep{raissi2017physics}. Physical phenomena are often represented through governing differential equations, which include initial and boundary conditions. These equations, comprising both linear and nonlinear partial differential equations (PDEs) and ordinary differential equations (ODEs), define the constraints that a solution must fulfill within a given domain. In the PINN approach, neural networks are employed to approximate the solutions of these differential equations based on the governing physical laws of a process, thus transforming the task of solving differential equations into an optimization problem~\citep{raissi2017physics, raissi2017physicspart2, raissi2019physics, karniadakis2021physics, Cuomo2022}.

Consider a system defined by a parameterized, nonlinear partial differential equation,
\begin{equation}
    u_t + \mathcal{N}[u; \bm{\lambda}] = 0, \quad x \in \Omega, \quad t \in [0, T],
    \label{eq:nonlinear_operator}
\end{equation}
where \( u(t, x) \) models the unknown (hidden) solution, \( u_t \) is the partial time derivative of \( u \), \( \mathcal{N} [\cdot; \bm{\lambda}] \) represents a differential operator dependent on the parameter \( \bm{\lambda} \), and $\Omega \subseteq \mathbb{R}^D$. This definition of partial differential equations covers a broad range of problems in physics, including kinetic equations and conservation laws~\citep{raissi2019physics}.

Given noisy measurements of the system, the objective is to tackle two primary problems. The first problem involves deriving solutions for differential equations from data~\citep{raissi2017physics}. This task, known as the \emph{forward problem}, aims to determine the hidden state \( u(t, x) \) of the system with known operator parameters \( \bm{\lambda} \)~\citep{raissi2017physics, raissi2019physics}. 
The second challenge involves the data-driven discovery of differential equations~\citep{Rudy2016DatadrivenDO, RaissiHidden, raissi2017physicspart2, raissi2019physics}. It seeks to determine the optimal parameters \( \bm{\lambda} \) that most accurately represent the system, based on limited and possibly noisy observations of the latent state \( u(t, x) \), a task known as the \emph{inverse problem}.

Let \( f(t, x) \) be defined as the left-hand side of Equation~\eqref{eq:nonlinear_operator},
\begin{equation}
    f := u_t + \mathcal{N}[u; \bm{\lambda}],
    \label{eq:physics_equation}
\end{equation}
and let \( u(t,x) \) be modeled by a deep neural network with parameters \( \bm{\theta} \). Using the neural network approximation in conjunction with the residual \( f(t,x) \) from Equation~\eqref{eq:physics_equation}, a physics-informed neural network can be developed that shares the same parameters as the network \( u(t,x) \)~\citep{raissi2017physicspart2}. Note that the parameters \( \bm{\lambda} \) of the differential operator are now integrated as parameters within the physics-informed model~\citep{raissi2019physics}. The required derivative terms for Equation~\eqref{eq:physics_equation} are computed using automatic differentiation~\citep{Baydin2018Vancouver}.

\begin{figure}[h]
  \centering
  \includegraphics[width=\linewidth]{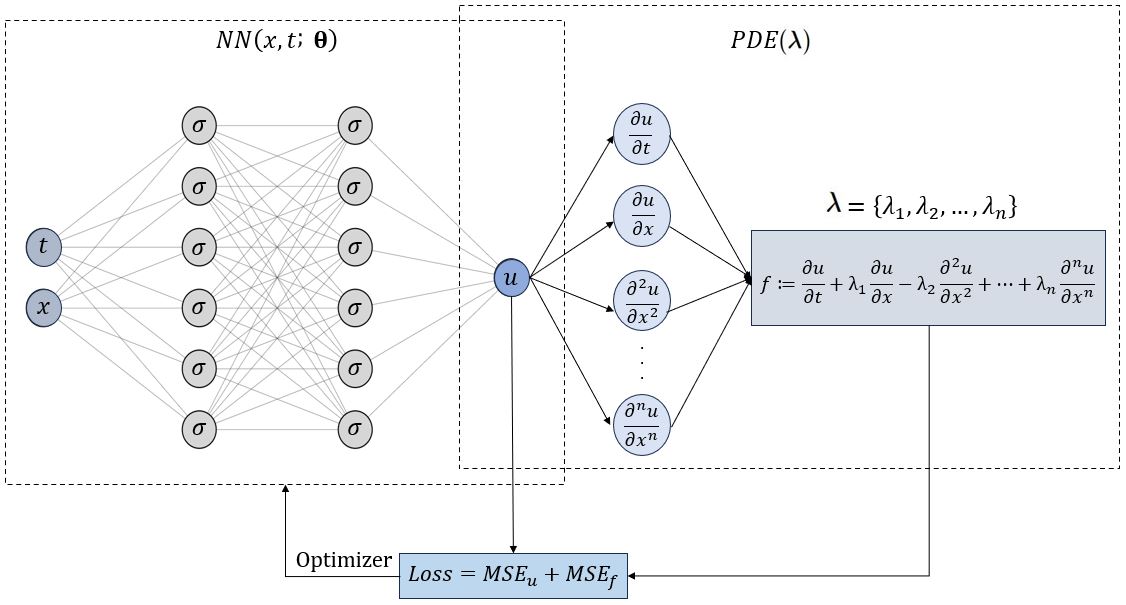}
  \caption{Illustration of a physics-informed neural network: A feed-forward neural network \( NN(x,t; \boldsymbol{\theta}) \) receives spatial and temporal coordinates, \( x \) and \( t \), as inputs to approximate the solutions of a physical system \(u\). Automatic differentiation is utilized to calculate the partial derivatives of \(u\) with respect to these inputs, which are then employed to formulate the loss function terms. By optimizing the loss function, the network simultaneously learns both its own parameters \( \boldsymbol{\theta} \) and the unknown parameters of the differential equation \( \bm{\lambda} \). In this schematic, the differential operator \( \mathcal{N}[\cdot] \) is chosen to be \( \lambda_1 \frac{\partial \hat{u}}{\partial x} - \lambda_2 \frac{\partial^2 \hat{u}}{\partial x^2} + \ldots + \lambda_n \frac{\partial^n \hat{u}}{\partial x^n} \).}
  \label{fig:PINNs}
\end{figure}

Figure~\ref{fig:PINNs} depicts an example of PINN architecture with a fully-connected feed-forward neural network with two hidden layers. The network takes as inputs the spatio-temporal variables, \(x\) and \(t\), and outputs the PDE solution \(u\). In this schematic, physical laws are integrated externally into the neural network as penalty losses to weakly enforce the physics constraints on the output. Additionally, the underlying physics can also be incorporated by embedding physical principles directly within the network's architecture or through the use of data augmentation strategies~\citep{karniadakis2021physics}. In forward problems, the PDE parameters \( \bm{\lambda} \) are predetermined, while in inverse problems, they are considered as learnable parameters.

Parameter optimization for the neural network \( u(t, x) \) and the unknown parameters of the differential equation is achieved through the minimization of the mean squared error loss~\citep{raissi2017physicspart2}, expressed as 
\begin{equation} 
    \mathrm{L} =  \mathrm{MSE}_{u} + \mathrm{MSE}_{f},
    \label{eq:MSE}
\end{equation}
where
\begin{align}
    \mathrm{MSE}_{u} &= \frac{1}{N} \sum_{i=1}^{N_u} \left| u(t^{i}_{u}, x^{i}_{u}) - u^{i} \right|^2, \label{eq:MSEu} \\
    \mathrm{MSE}_{f} &= \frac{1}{N} \sum_{i=1}^{N_f} \left| f(t^{i}_{f}, x^{i}_{f}) \right|^2, \label{eq:MSEf}
\end{align}
in which \( \{t_u^{i}, x_u^{i}, u^{i}\}_{i=1}^{N_{u}} \) represents the training data, including the initial condition and boundary condition for \( u(t, x) \), while \( \{t_f^{i}, x_f^{i}\}_{i=1}^{N_{f}} \) denote the collocation points for \( f(t, x) \). The term \( \mathrm{MSE}_{u} \) ensures that the model's predictions align with the observed data for \( u(t, x) \), and \( \mathrm{MSE}_{f} \) penalizes deviations from the solution of the differential equation at the selected collocation points~\citep{raissi2017physicspart2}. 

Empirical studies show that the integration of physical laws into deep learning architectures through PINNs acts as a regularization mechanism and can yield good predictive accuracy~\citep{karniadakis2021physics, raissi2017physicspart2}. This is particularly true when the underlying differential equation has a unique solution and is well-posed. The success of this approach also depends on having a sufficient number of collocation points and an adequately expressive neural network~\citep{mishra2021estimates}. This approach integrates physical laws into deep learning architectures, effectively constraining the model to a lower-dimensional space, which allows training with smaller data sets~\citep{raissi2019physics}. In this context, neural networks may be configured to respect symmetries, invariances, or conservation laws present within a physical system that correlate with the patterns observed in the training data~\citep{raissi2017physics}. 

\section{Material and Methods}
\label{app:method}

This section is divided into two parts: data preprocessing and mathematical modeling of froth flotation. Section \ref{app:Prepro} discusses the data preprocessing approach, focusing on outlier removal, which is applied to the simulated datasets to prepare them for training purposes. Section \ref{app:model} details three different mathematical models that represent the complex dynamics of froth flotation processes and are employed to construct physics-informed neural networks.
%

\subsection{Data Preprocessing}
\label{app:Prepro}

In this study, the analysis focused on datasets provided by Metso Oyj for two rougher froth flotation cells, which are part of a larger operational circuit in a gold flotation process. This circuit includes various stages, with the rougher sub-process itself comprising four flotation cells. Among these, two cells were selected for detailed analysis. For each flotation cell, three datasets were generated from a digital twin model that simulates the gold flotation circuit by dynamically integrating real-process data with physics-based models calibrated using historical data~\citep{Metso2024Geminex}. The same digital twin was used in a study by~\citet{Zeb2024}, where the authors analysed cleaner cells. These datasets include 14 different variables as shown in Table \ref{table:flotation_cell_variables}, with the first twelve variables serving as inputs to all the machine learning models discussed in this paper, while the gold grades in the concentrate and tailings serve as the outputs. The data, collected over various periods within a span of nearly half a year at a consistent five-minute sampling rate, reflect the dynamic and evolving conditions of the actual plant operations and include noise and outliers.

\begin{table}[h!]
\captionsetup{justification=raggedright,singlelinecheck=false}
\caption{Variables in the datasets of froth flotation cells.}

\label{table:flotation_cell_variables}
\renewcommand{\arraystretch}{1.3} 
\noindent 
\begin{tabular}{l@{\hspace{0.7in}}l@{\hspace{0.4in}}l}
\hline
\textbf{Definition} & \textbf{Notation} & \textbf{Unit} \\
\hline
Time & \( t \) & \( min \) \\
Air flow rate fed to the cell & \( Q_\textrm{air} \) & \( m^3\cdot h^{-1} \) \\
Level of materials in the cell & \( h \) & \(\% \) \\
Proportion of solids in the feed & \( C_{s} \) & \(\% \) \\
Solids recovery rate in the feed & \( R_{s,\textrm{feed}} \) & \(\% \) \\
Gold grade in the feed & \( C_\textrm{feed} \) & \( g\cdot t^{-1} \) \\
Gold recovery rate in the feed & \( R_{\textrm{Au},\textrm{feed}} \) & \(\% \) \\
P80 particle size & \( P_{80} \) & \( \mu m \) \\
Pulp volumetric flow in the feed & \( Q_\textrm{feed} \) & \( m^3\cdot h^{-1} \) \\
Total solids flow in the feed & \( F_{s,\textrm{feed}} \) & \( t\cdot h^{-1} \) \\
Pulp volumetric flow in the tailings & \( Q_t \) & \( m^3\cdot h^{-1} \) \\
Pulp volumetric flow in the concentrate & \( Q_c \) & \( m^3\cdot h^{-1} \) \\
Gold grade in the tailings & \( C_p \) & \( g\cdot t^{-1} \) \\
Gold grade in the concentrate & \( C_f \) & \( g\cdot t^{-1} \) \\
\hline
\end{tabular}
\end{table}

\begin{figure}[h]
  \centering
  \includegraphics[width=0.82\linewidth]{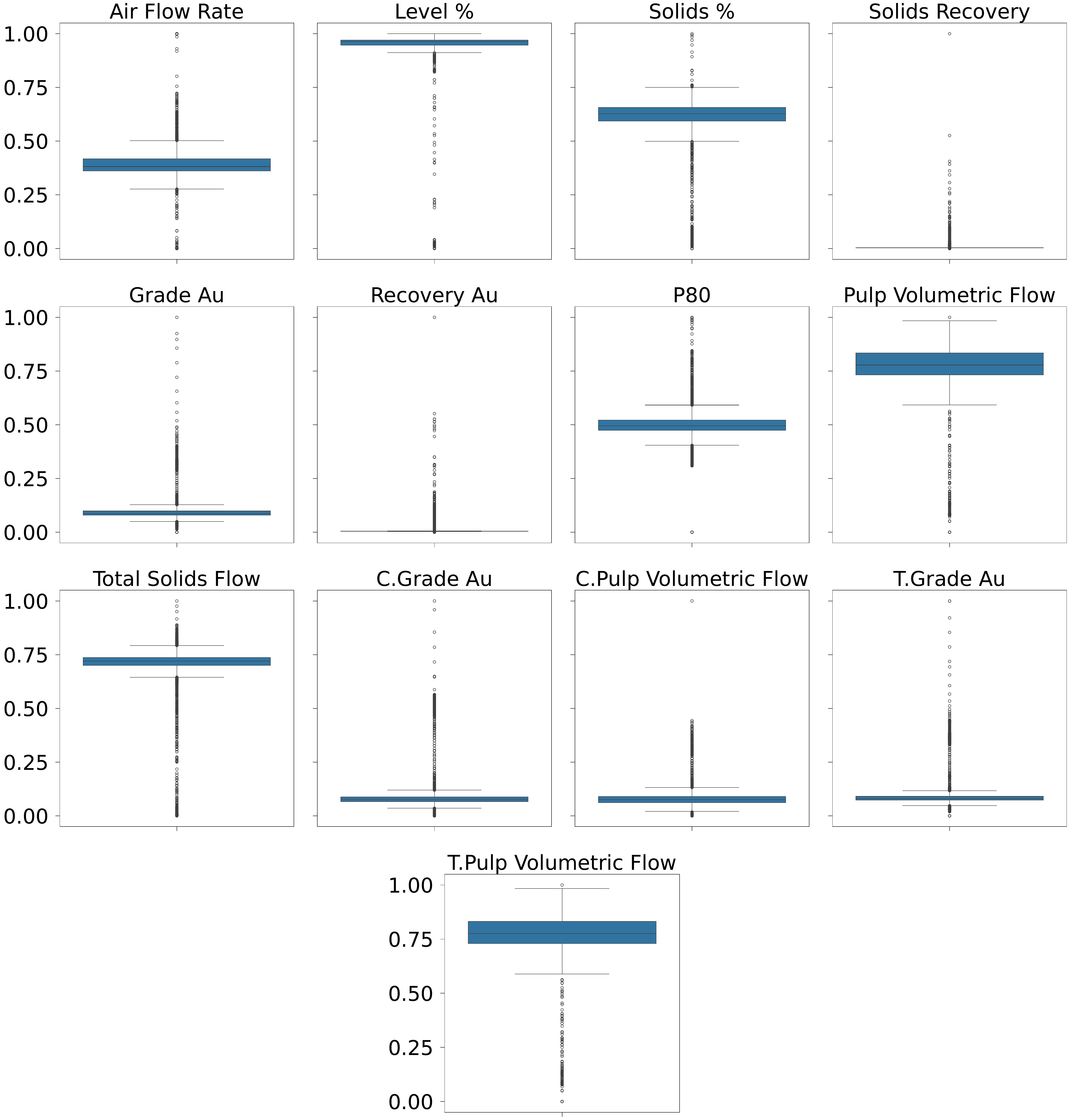}
  \caption{Normalized box plots displaying data point distributions and identifying the potential outliers in one of the datasets. Here, 'C' stands for concentrate and 'T' stands for tail.}
  \label{fig:boxplot}
\end{figure}

To manage outliers in the datasets, the interquartile range (IQR) technique was applied. The IQR, a measure of statistical dispersion in descriptive statistics, quantifies the variability within a dataset and is used for identifying the least contributing and extreme data points~\citep{Dekking2005ProbabilityStatistics, Kaltenbach2012Statistics}. In this approach, the dataset is divided into four quartiles using linear interpolation, with Q1 and Q3 defined as the 25th and 75th percentiles, respectively, and the IQR is then calculated as the difference between Q3 and Q1~\citep{Kaltenbach2012Statistics, Dekking2005ProbabilityStatistics, Kokoska2000Statistics}. Outliers are then defined as data points falling below \( Q1 - 1.5 \times \text{IQR} \) or above \( Q3 + 1.5 \times \text{IQR} \). Based on domain expertise, outliers in the datasets were identified as errors from simulations and, therefore, were removed to enhance the reliability of the datasets for further analysis. To visually represent the distribution of the datasets and identify outliers, box plots~\citep{tukey1997exploratory} were used, as depicted in Figure~\ref{fig:boxplot} for one of the datasets. To retain confidentiality, all variables shown in the figures in this paper have been scaled to a range between 0 and 1.

\subsection{Froth Flotation Models}
\label{app:model}

In this section, the mathematical models used in the learning algorithms to model froth flotation dynamics are discussed in detail. Froth flotation demonstrates complex, nonlinear, and non-stationary dynamics affected by a multitude of variables and micro-process interactions, posing significant challenges for precise modeling and prediction~\citep{Putz2015HybridMPC}. In this study, three distinct dynamic models based on first-principles, specifically kinetic and mass balance laws, are employed to develop physics-informed neural networks. These networks are designed to predict the gold grade in the concentrate of flotation cells. The approach integrates physics-based mathematical modeling with data-driven machine learning to enhance the generalization capabilities in different operating conditions encountered in flotation cells.

\subsubsection{Bidirectional Material Transfer Dynamics Model}

\begin{figure}[h]
  \centering
  \includegraphics[width=0.45\linewidth]{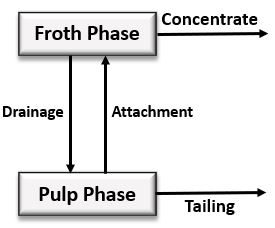}
  \caption{Material transition in a flotation process across froth and pulp phases (adapted from \citet{lynch1981mineral}).}
  \label{fig:flotation_phase}
\end{figure}

In froth flotation, kinetic models are developed by associating the process dynamics with chemical reactions to effectively capture the behavior of flotation~\citep{Quintanilla2021a}. These models can be formulated to include material transfers not only from the pulp to the froth phase but also vice versa, thereby capturing the entirety of material movement. Figure~\ref{fig:flotation_phase} depicts a schematic representation of particle movement within a flotation cell, illustrating the transition of particles between the pulp and froth phases. The diagram demonstrates how particles transfer from the pulp phase due to non-selective entrainment or selective attachment, along with the return movement of particles from the froth phase back to the pulp, driven by drainage.

As discussed in~\citet{Putz2015HybridMPC}, the dynamics of a rougher flotation cell can be formulated by considering the interactions between the froth and pulp phases. This model utilizes coupled dynamic equations, derived through mass balances in each phase, accounting for the attachment and drainage flows between them. The assumptions underlying the model in~\citet{Putz2015HybridMPC} include perfect mixing within each phase, a constant air flow rate into the cell, and material transfers due to collection and drainage rates, with the flotation cell having a constant horizontal cross-sectional area. However, the model presented in this paper takes into account variations in air flow rate into the cell \( Q_\textrm{air} \), which affects the drainage and collection flows between phases. The dynamic equations for the pulp and froth masses in a single flotation cell and specific to a mineral class are defined as
%
\begin{align}
    \frac{dm_p}{dt} &= M_\textrm{feed} + \alpha_f Q_\textrm{air} m_f - (\alpha_p Q_\textrm{air} + \frac{Q_t}{V_p}) m_p, \label{eq:coupled_pulp} \\
    \frac{dm_f}{dt} &= \alpha_p Q_\textrm{air} m_p - (\alpha_f Q_\textrm{air} + \frac{Q_c}{V_f}) m_f, \label{eq:coupled_froth}
\end{align}
respectively, where \( M_\textrm{feed} \) denotes the mass flow in the feed. The variables \( V_p \) and \( V_f \) represent the volumes of the pulp and froth phases respectively, while \( m_p \) and \( m_f \) indicate the masses in each phase. In the above equations, the collection and drainage coefficients are denoted by \( \alpha_p \) and \( \alpha_f \), respectively. Note that in all the mathematical models discussed in this paper, the volumetric flow rates of the tail \( Q_t \) and the concentrate \( Q_c \) are assumed to be known, as they can be measured on-line.

Further expressing the feed mass flow, froth mass, and pulp mass in terms of their concentrations, we have
\begin{align}
M_{\text{feed}} &= C_{\text{feed}} Q_{\text{feed}}, \label{eq:M_feed} \\
m_f &= C_f V_f, \label{eq:m_f} \\
m_p &= C_p V_p, \label{eq:m_p}
\end{align}
and assuming constant volumes \( V_p \) and \( V_f \), the coupled Equations~\eqref{eq:coupled_pulp} and~\eqref{eq:coupled_froth} can be reformulated in terms of concentrations as:
\begin{align}
    \frac{dC_p}{dt} &= C_\textrm{feed} \frac{Q_\textrm{feed}}{V_p} + \alpha_f Q_\textrm{air} C_f \frac{V_f}{V_p} - (\alpha_p Q_\textrm{air} + \frac{Q_t}{V_p}) C_p, \label{eq:coupled_pulp_1} \\
    \frac{dC_f}{dt} &= \alpha_p Q_\textrm{air} C_p \frac{V_p}{V_f} - (\alpha_f Q_\textrm{air} + \frac{Q_c}{V_f}) C_f. \label{eq:coupled_froth_1}
\end{align}

The functions \( f_{C_f}(t, \mathbf{x}) \) and \( f_{C_p}(t, \mathbf{x}) \) can now be defined, with the unknown parameters $V_p$, $\alpha_f$, $V_f$, and $\alpha_p$ being represented as $\lambda_1$, $\lambda_2$, $\lambda_3$, and $\lambda_4$, respectively, to be given by
\begin{align}
    f_{C_p} &\coloneq \frac{dC_p}{dt} - C_\textrm{feed} \frac{Q_\textrm{feed}}{\lambda_1} - \lambda_2 Q_\textrm{air} C_f \frac{\lambda_3}{\lambda_1} + (\lambda_4 Q_\textrm{air} + \frac{Q_t}{\lambda_1}) C_p, \label{eq:equation_cp} \\
    f_{C_f} &\coloneq \frac{dC_f}{dt} - \lambda_4 Q_\textrm{air} C_p \frac{\lambda_1}{\lambda_3} + (\lambda_2 Q_\textrm{air} + \frac{Q_c}{\lambda_3}) C_f. \label{eq:equation_cf}
\end{align}

Here, the target variables \( \mathbf{u} = (C_p, C_f) \) are approximated by a deep neural network \( \mathbf{u}(t, \mathbf{x}) \), where \( \mathbf{x} \) comprises variables two through twelve as listed in Table \ref{table:flotation_cell_variables}. This network, along with Equations~\eqref{eq:equation_cp} and~\eqref{eq:equation_cf}, is used to form the physics-informed neural network. Here, the learnable parameters of the differential equations are denoted by \( \boldsymbol{\lambda} = (\lambda_1, \lambda_2, \lambda_3, \lambda_4) \). Automatic differentiation~\citep{Baydin2018Vancouver} is then employed to compute the necessary partial derivatives of the deep network outputs with respect to \( t \), which are used in the loss function. The optimization of the shared parameters in the neural network \( \mathbf{u}(t, \mathbf{x}) \), together with the parameters of the differential equations \( \boldsymbol{\lambda} \), is achieved through the minimization of the mean squared error loss 
\begin{equation} 
    \mathrm{L} = \mathrm{MSE}_\mathbf{u} + \mathrm{MSE}_\mathbf{f},
    \label{eq:MSE_PIIN1}
\end{equation}
where 
\begin{align}
    \mathrm{MSE}_\mathbf{u} &= \frac{1}{N} \sum_i^N \left\| \mathbf{u}(t^{i}, \mathbf{x}^{i}) - \mathbf{u}^{i} \right\|^2, \label{eq:MSEu_PIIN1} \\
     \mathrm{MSE}_\mathbf{f} &= \frac{1}{N} \sum_i^N \left\| \mathbf{f}(t^{i}, \mathbf{x}^{i}) \right\|^2 = \frac{1}{N} \sum_i^N \left( \left| f_{C_p}(t^{i}, \mathbf{x}^{i})\right|^2 + \left|f_{C_f}(t^{i}, \mathbf{x}^{i}) \right|^2 \right). \label{eq:MSEf_PIIN1}
\end{align}

In the equations above, \( \{t^{i}, \mathbf{x}^{i}, \mathbf{u}^{i}\}_{i=1}^N \) denote the training data points for \( \mathbf{u}(t, \mathbf{x}) \) and \( \mathbf{f} \) consists of components \( f_{C_p} \) and \( f_{C_f} \).


%
%
%

\subsubsection{Unidirectional Material Transfer Dynamics Model}

\citet{Cubillos1997FlotationControl} developed an alternative kinetic model that considers material transfer solely from the pulp to froth phase to model flotation dynamics. The dynamic behavior of a single flotation cell for a specific mineral class is represented using mass conservation equations for the pulp and froth phases as
\begin{align}
    \frac{dm_{p}}{dt} &= M_\textrm{feed} - \left(\frac{Q_t}{V_p}\right) m_p - R, \label{eq:model2_pulp} \\
    \frac{dm_{f}}{dt} &= - \left(\frac{Q_c}{V_f}\right) m_f + R, \label{eq:model2_froth}
\end{align}
where \( M_\textrm{feed} \) indicates the mass flow in the feed, the volumes of the pulp and froth phases are denoted by \( V_p \) and \( V_f \) respectively, \( m_p \) and \( m_f \) are the masses within each phase, and \( R \) represents the average rate of flotation. 

With the assumption that the volumes \( V_p \) and \( V_f \) remain constant, and by utilizing Equations~\eqref{eq:M_feed}, \eqref{eq:m_f}, and (\ref{eq:m_p}), the conservation Equations~\eqref{eq:model2_pulp} and \eqref{eq:model2_froth} can be rewritten in terms of concentrations as:
\begin{align}
    \frac{dC_p}{dt} &= C_\textrm{feed} \frac{Q_\textrm{feed}}{V_p} - \frac{Q_t}{V_p} C_p - \frac{R}{V_p}, \label{eq:coupled_pulp_2} \\
    \frac{dC_f}{dt} &= - \frac{Q_c}{V_f} C_f + \frac{R}{V_f}. \label{eq:coupled_froth_2}
\end{align}

The functions \( f_{C_f}(t, \mathbf{x}) \) and \( f_{C_p}(t, \mathbf{x}) \) can then be defined by the equations below, where the unknown volumes \( V_p \) and \( V_f \) are represented by learnable parameters \( \lambda_1 \) and \( \lambda_2 \) respectively.

\begin{align}
    f_{C_p} &\coloneq \frac{dC_p}{dt} - C_\textrm{feed} \frac{Q_\textrm{feed}}{\lambda_1} + \frac{Q_t}{\lambda_1} C_p + \frac{R}{\lambda_1} \label{eq:model2_fcp} \\
    f_{C_f} &\coloneq \frac{dC_f}{dt} + \frac{Q_c}{\lambda_2} C_f - \frac{R}{\lambda_2} \label{eq:model2_fcf}
\end{align}

Then, the approximation of the target variables \( \mathbf{u} = (C_p, C_f) \) is achieved using a deep feed-forward neural network \( \mathbf{u}(t, \mathbf{x}) \). It should be noted that the average flotation rate \( R \) is unknown and is estimated through a shallow neural network \( R(t, \mathbf{x}, \mathbf{u}(t, \mathbf{x})) \). These networks along with Equations~\eqref{eq:model2_fcp} and~\eqref{eq:model2_fcf} are employed to form a physics-informed neural network, where the unknown parameters of the differential equations are denoted by \( \boldsymbol{\lambda} = (\lambda_1, \lambda_2) \). The network \( \mathrm{R}(t, \mathbf{x}, \mathbf{u}(t, \mathbf{x})) \) is not designed to directly map inputs to a known output; rather, it takes in both spatio-temporal inputs and the outputs approximated from \( \mathbf{u}(t, \mathbf{x}) \) to estimate the average rate of flotation \( R \). The shared parameters of \( \mathbf{u}(t, \mathbf{x}) \) and \( \mathrm{R}(t, \mathbf{x}, \mathbf{u}(t, \mathbf{x})) \), along with the unknown parameters \( \boldsymbol{\lambda} \), are optimized by minimizing the mean squared error, as specified in Equations~\eqref{eq:MSE_PIIN1}, \eqref{eq:MSEu_PIIN1}, and~\eqref{eq:MSEf_PIIN1}.
%
%
%
%

\subsubsection{Overall Mass Balance Model}

The final model discussed in this paper addresses the total mass balance of a mineral within a flotation cell. This model captures the overall mass conservation by incorporating the mass flows of the feed, concentrate, and tail within the system, as expressed by~\citep{lynch1981mineral}
\begin{equation}
\frac{dm}{dt} = M_\textrm{feed} - M_c - M_t,
\label{eq:total_mass_balance}
\end{equation}
where \( m \) denotes the total mineral mass within the flotation cell, \( M_\textrm{feed} \) refers to the rate of mass entering as feed, while \( M_c \) and \( M_t \) represent the mass flow rates of the concentrate and tailings, respectively. The total mass \( m \) can be represented by the sum of the masses \( m_f \) and \( m_p \). Considering the on-line measurements of volumetric flow rates of the tail and the concentrate, and following the approach outlined in Equations~\eqref{eq:M_feed}, \eqref{eq:m_f}, and~\eqref{eq:m_p}, the total mass balance equation can be reformulated in terms of concentrations as
%
\begin{equation}
\frac{d\left( C_f V_f + C_p V_p \right)}{dt} = Q_\textrm{feed} C_\textrm{feed} - Q_c C_f - Q_t C_p,
\label{eq:mass_balance}
\end{equation}
in which \( C_\textrm{feed} \) refers to the feed concentration, while froth and pulp concentrations are denoted by \( C_f \) and \( C_p \), respectively. The variables \( Q_{\text{feed}} \), \( Q_c \), and \( Q_t \) denote the volumetric flow rates of the feed, concentrate, and tailings, respectively. Additionally, the volumes of the froth and pulp phases are denoted by \( V_f \) and \( V_p \).

Rearranging Equation~\eqref{eq:mass_balance} with the assumption that the volumes \( V_p \) and \( V_f \) remain constant, the dynamic equation for the rate of change in froth concentration can be expressed as
\begin{equation}
\frac{dC_f}{dt} = \frac{Q_\textrm{feed}}{V_f} C_\textrm{feed} - \frac{Q_c}{V_f} C_f - \frac{Q_t}{V_f} C_p - \frac{dC_p}{dt} \frac{V_p}{V_f}.
\label{eq:mass_balance_froth}
\end{equation}

Let \( f_{C_f}(t, \mathbf{x}) \) now be defined, where the unknown parameters \( \lambda_1 \) and \( \lambda_2 \) correspond to the volumes \( V_f \) and \( V_p \) respectively, as
\begin{equation}
f_{C_f} := \frac{dC_f}{dt} - \frac{Q_\textrm{feed}}{\lambda_1} C_\textrm{feed} + \frac{Q_c}{\lambda_1} C_f + \frac{Q_t}{\lambda_1} C_p + \frac{dC_p}{dt} \frac{\lambda_2}{\lambda_1},
\label{eq:phys_mass_balance_froth}
\end{equation}
that is used in conjunction with a deep neural network \( \mathbf{u}(t, \mathbf{x}) \), which approximates the variables \( \mathbf{u} = (C_p, C_f) \), to construct a physics-informed neural network. In the above equation, the differential equation is parameterized by \( \boldsymbol{\lambda} = (\lambda_1, \lambda_2) \). Finally, the optimization is achieved by minimizing the mean squared error loss
%
\begin{equation} 
    \mathrm{L} = \mathrm{MSE}_\mathbf{u} + \mathrm{MSE}_f,
    \label{eq:equation_53}
\end{equation}
where 
\begin{align}
    \mathrm{MSE}_\mathbf{u} &= \frac{1}{N} \sum_i^N \left\| \mathbf{u}(t^{i}, \mathbf{x}^{i}) - \mathbf{u}^{i} \right\|^2, \label{eq:equation_54} \\
    \mathrm{MSE}_f &= \frac{1}{N} \sum_i^N \left| f(t^{i}, \mathbf{x}^{i}) \right|^2, \label{eq:equation_55}
\end{align}
and \( f \) corresponds to \( f_{C_f} \).

\section{Results}
\label{app:result}

In this section, the performance of the proposed physics-informed neural networks is evaluated. The study utilized simulated data from two rougher flotation cells, each containing three distinct datasets collected over different periods of time. For each cell, the models were separately trained, evaluated, and tested. The analysis begins with the results for Cell I, followed by those for Cell II. The performance of the physics-informed neural networks is compared with that of purely data-driven machine learning models, demonstrating the strengths of the physics-informed methodology in more accurately capturing the froth flotation dynamics.

\subsection{Predictive Analysis for Flotation Cell I}

\begin{figure}[h]
  \centering
  
  \begin{subfigure}{\linewidth}
    \centering
    \includegraphics[width=0.95\linewidth]{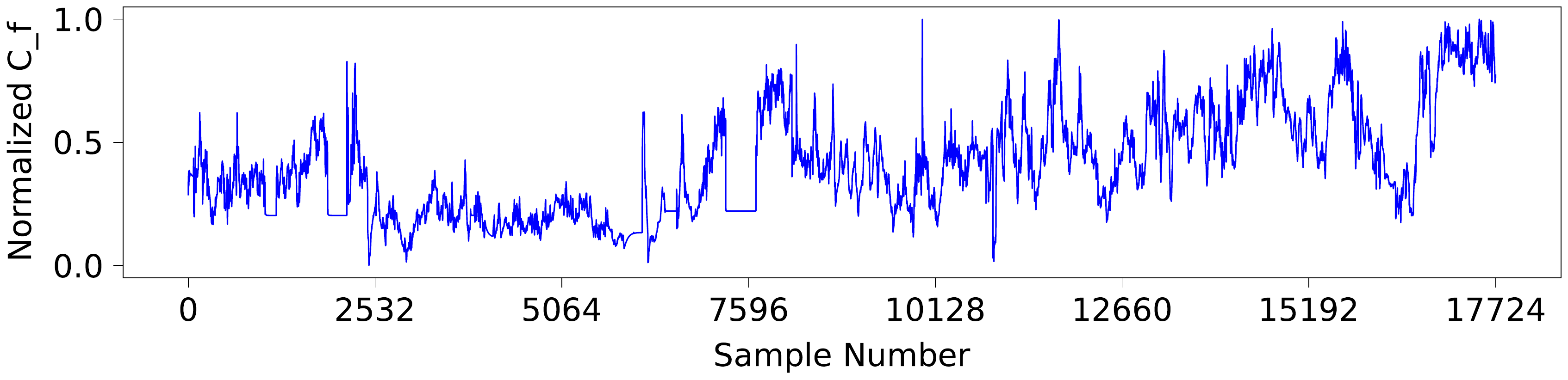}
    \caption{Training dataset}
    \label{fig:train_dataset}
  \end{subfigure}%
  \vspace{5mm} 

  \begin{subfigure}{\linewidth}
    \centering
    \includegraphics[width=0.95\linewidth]{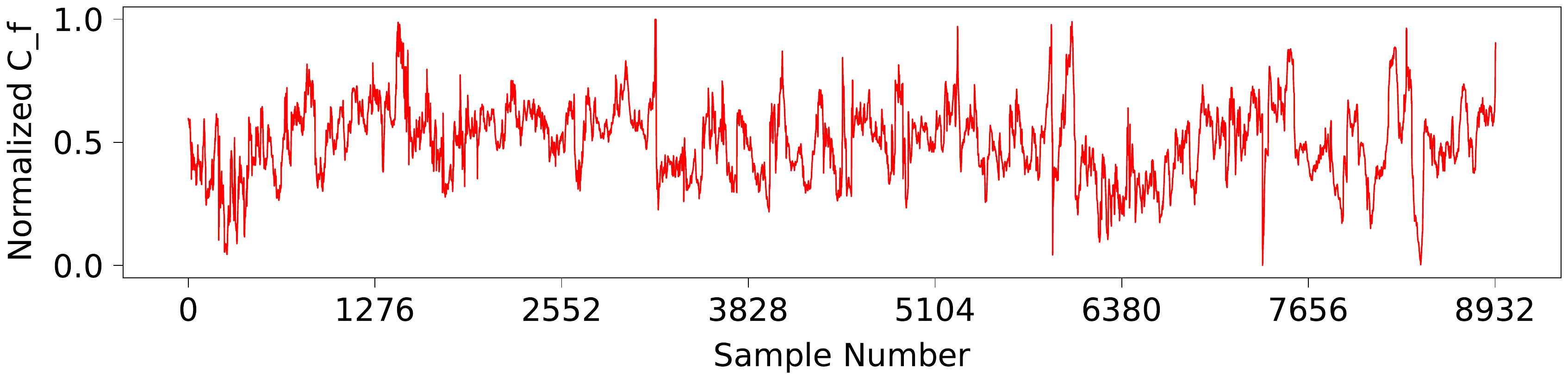}
    \caption{Validation dataset}
    \label{fig:validation_dataset}
  \end{subfigure}%
  \vspace{5mm} 

  \begin{subfigure}{\linewidth}
    \centering
    \includegraphics[width=0.95\linewidth]{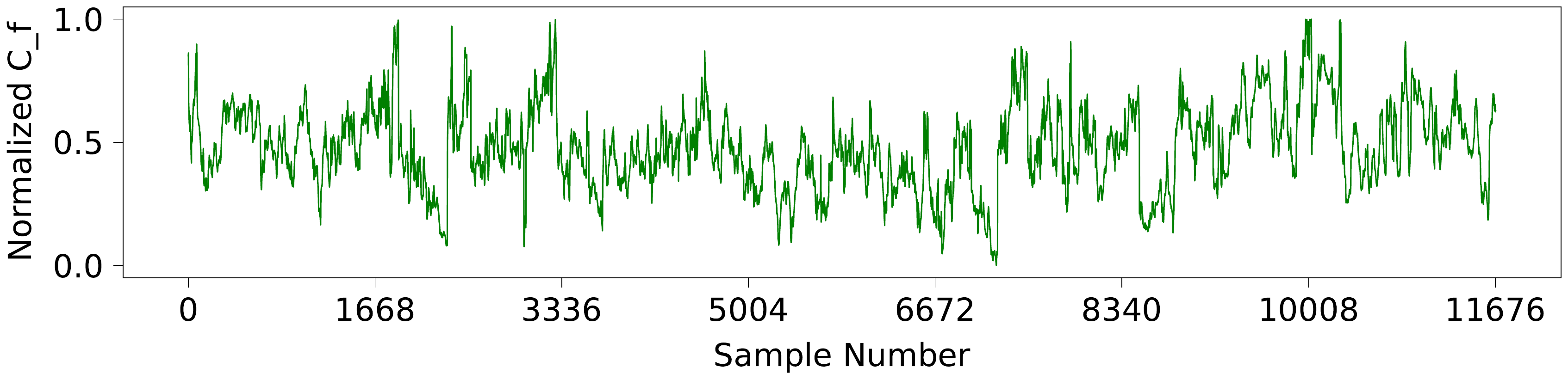}
    \caption{Test dataset}
    \label{fig:test_dataset}
  \end{subfigure}

  \caption{Distribution patterns of concentrate gold grade (\( C_f \)) across training, validation, and test datasets for Cell I.}
  \label{fig:data_sets_cell3}
\end{figure}

For Cell I, the analysis was conducted using three separate datasets for training, validation, and testing of the models. Specifically, the training dataset included 17724 data points, the validation set comprised 8936 data points, and the test set contained 11679 data points. These data points, related to the variables listed in Table~\ref{table:flotation_cell_variables}, are distributed across the spatio-temporal domain. Figure~\ref{fig:data_sets_cell3} depicts how the concentrate gold grade (\( C_f \)), the variable of interest, is distributed within these datasets.

%
%
%
%
%
%

In this research, purely data-driven and physics-informed neural networks, as well as traditional machine learning models such as linear regression, random forest, and decision tree, were employed for predicting the variable of interest, gold grade in the concentrate (\(C_f\)) of the froth flotation cells. The neural network models utilize a multilayer perceptron architecture \( \mathbf{u}(t, \mathbf{x}) \) with three hidden layers containing 256, 512, and 256 neurons, respectively, and use the hyperbolic tangent activation function. All machine learning models were designed to use the first twelve variables listed in Table~\ref{table:flotation_cell_variables} as inputs, and to output both the gold grade in the concentrate (\(C_f\)) and in the tailings (\(C_p\)). The optimization of model parameters was achieved by minimizing the mean squared error, as outlined in Section \ref{app:method} for the PINNs and by applying \( \mathrm{MSE}_\mathbf{u} \) for the data-driven models. The optimization process employed the Adam algorithm~\citep{Kingma2014AdamAM}, a gradient descent-based optimizer, with a learning rate set at \( 10^{-5} \). The hyperparameters of all the machine learning models were calibrated during the training phase, based on optimization results, to ensure accurate predictions of the variable of interest \( C_p \). The implementation of the models was carried out using Python~\citep{van1995python} with neural networks implemented in the PyTorch framework~\citep{paszke2019pytorch} and other machine learning models developed using the Scikit-learn library~\citep{scikit-learn}.

On the other hand, the auxiliary neural network \( R(t,\mathbf{x}, \mathbf{u}(t,\mathbf{x})) \), used in the unidirectional material transfer dynamics model to estimate the average flotation rate \( R \), takes the spatio-temporal variables \(x\) and \(t\), along with outputs estimated by \( \mathbf{u}(t, \mathbf{x}) \), as inputs, with \(x\) being variables two to twelve as outlined in Table \ref{table:flotation_cell_variables}. This network is designed as a shallow network with a single hidden layer containing 100 neurons. Otherwise, its architectural design and hyperparameters are the same as the primary neural network discussed earlier. The parameters of this auxiliary network, together with \( \mathbf{u}(t, \mathbf{x}) \) are collectively optimized through a unified optimization algorithm. The hyperparameters of the neural networks, the number of hidden layers and the respective number of neurons employed in modeling flotation Cell I are detailed in Table \ref{table:hyperparameter_cell3}.

\begin{table}[h]
\caption{Neural network configurations and hyperparameters for modeling flotation Cell I.}
\label{table:hyperparameter_cell3}
\renewcommand{\arraystretch}{1.3} 
\begin{tabular}{l@{\hspace{0.9in}}l}
\hline
\textbf{Hyperparameter} & \textbf{Configuration} \\
\hline
Optimizer & Adam \\
Activation function & Tanh \\
Learning rate & $10^{-5}$ \\
Batch size & $128$ \\
\hline
\multicolumn{2}{l}{\textbf{Network \( \mathbf{u}(t, \mathbf{x}) \) Architecture}} \\
\hline
Number of hidden layers & 3 \\
Number of neurons in the hidden layers & 256, 512, 256 \\
\hline
\multicolumn{2}{l}{\textbf{Network \( R(t,\mathbf{x}, \mathbf{u}(t,\mathbf{x})) \) Architecture}} \\
\hline
Number of hidden layer & 1 \\
Number of neurons in the hidden layer & 100 \\
\hline
\end{tabular}
\end{table}

Considering the domain knowledge and parameters specific to the flotation cells, hard constraints on volumes were explicitly imposed within the architecture of the neural network \( \mathbf{u}(t, \mathbf{x}) \) in all PINN models. Particularly, the total volume of a cell, which is the summation of the volumes of the froth and pulp phases, was set to 26.7. Moreover, the froth volume was limited to range between 4\% and 7\% of the total volume, whereas 93\% to 96\% of the total volume was designated for the volume of the pulp phase.

In this research, a deep feed-forward neural network with relatively simple architecture and no additional regularization methods such as dropout or L1/L2 penalties was employed. The models were trained using unnormalized data to be able to capture the solution of the physical laws; however, as already mentioned before, for confidentiality as required by Metso Oyj, all the variables in the datasets and the model predictions are scaled to a range between 0 and 1.

For each neural network model utilized in this study, an early stopping mechanism was applied, configured with a patience of 20000 and a tolerance threshold of \(10^{-5}\). The early stopping was not employed as a regularization method but rather to prevent excessively long training durations while ensuring the models reached optimal performance. Throughout the training and validation phases, the model achieving the lowest validation mean squared error was preserved. Finally, the performance of the selected models was assessed on the test dataset.

\begin{figure}[htbp]
  \centering
  
  \begin{subfigure}[b]{0.495\linewidth}
    \centering
    \includegraphics[width=\linewidth]{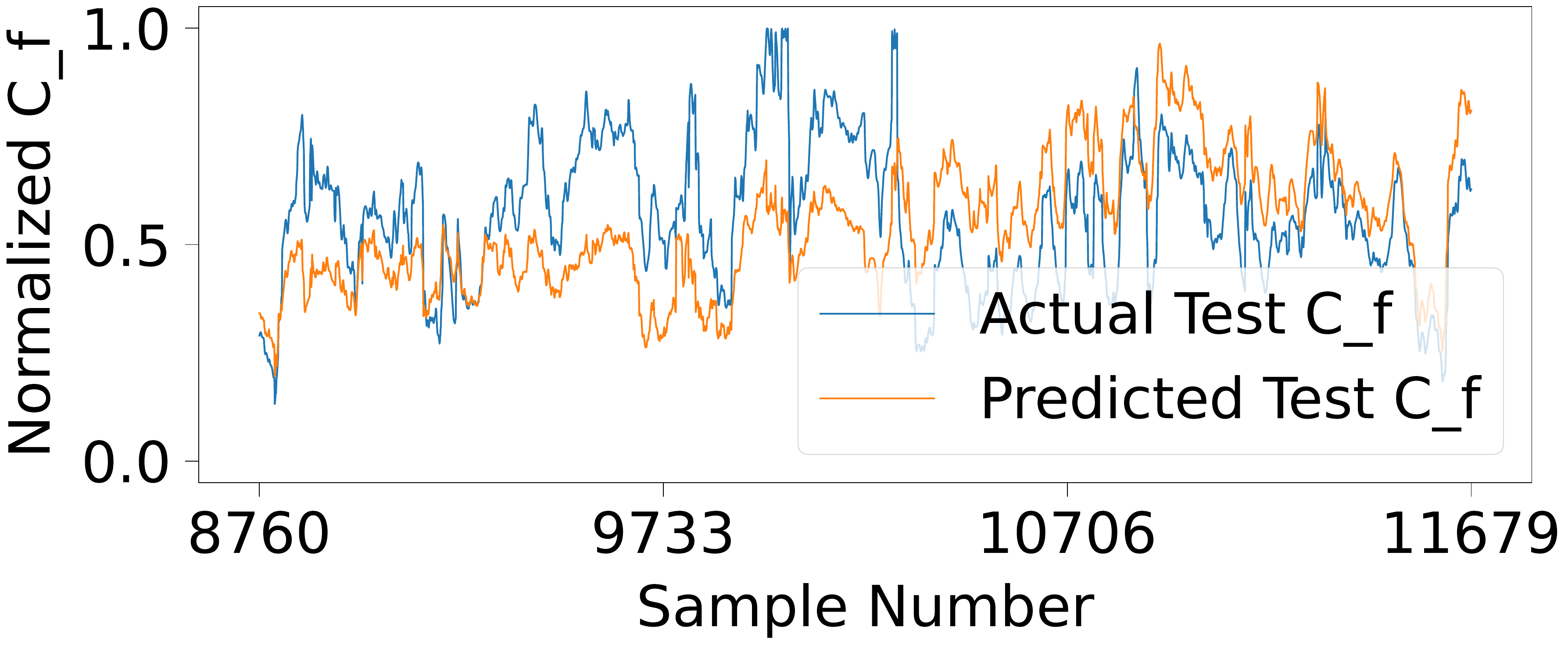}
    \caption{Linear regression}
    \label{fig:lr}
  \end{subfigure}
  \hfill 
  \begin{subfigure}[b]{0.495\linewidth}
    \centering
    \includegraphics[width=\linewidth]{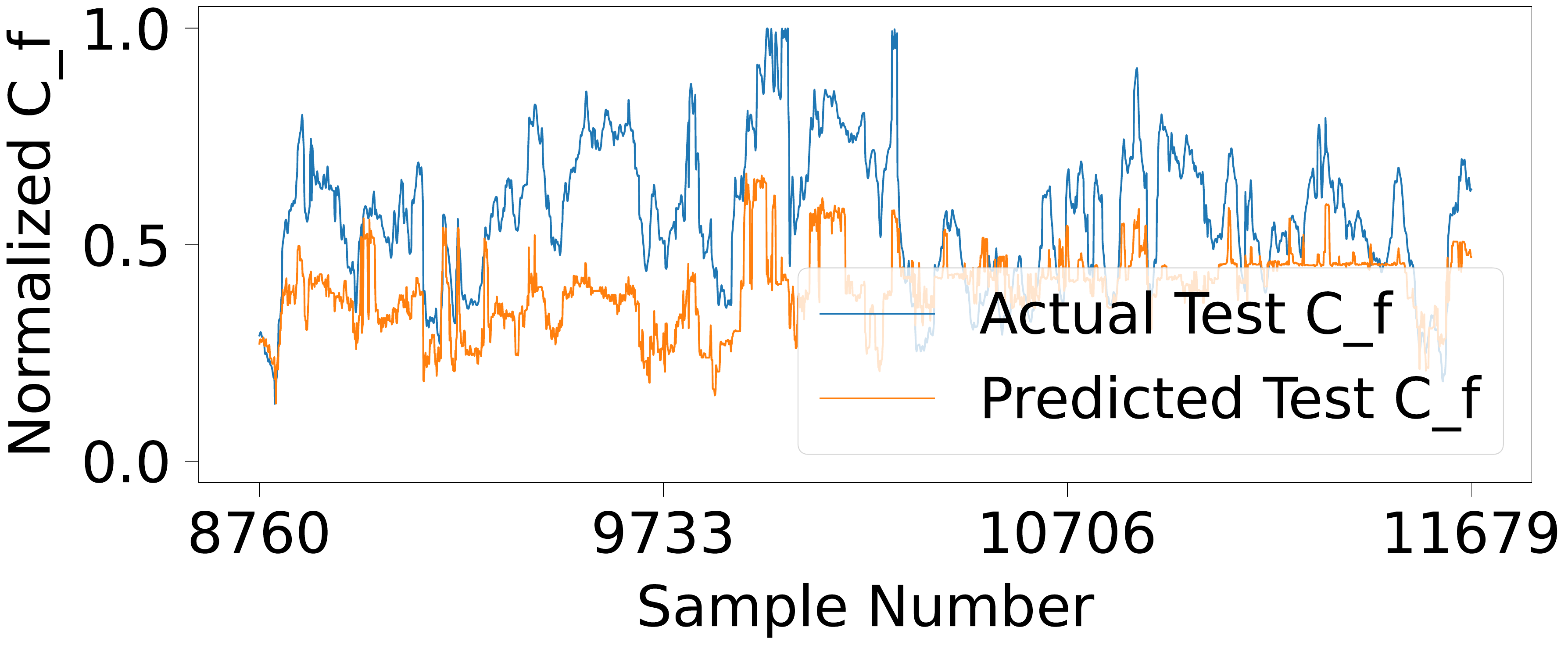}
    \caption{Random forest}
    \label{fig:rf}
  \end{subfigure}
  
  \vspace{5mm} 
  
  \begin{subfigure}[b]{0.495\linewidth}
    \centering
    \includegraphics[width=\linewidth]{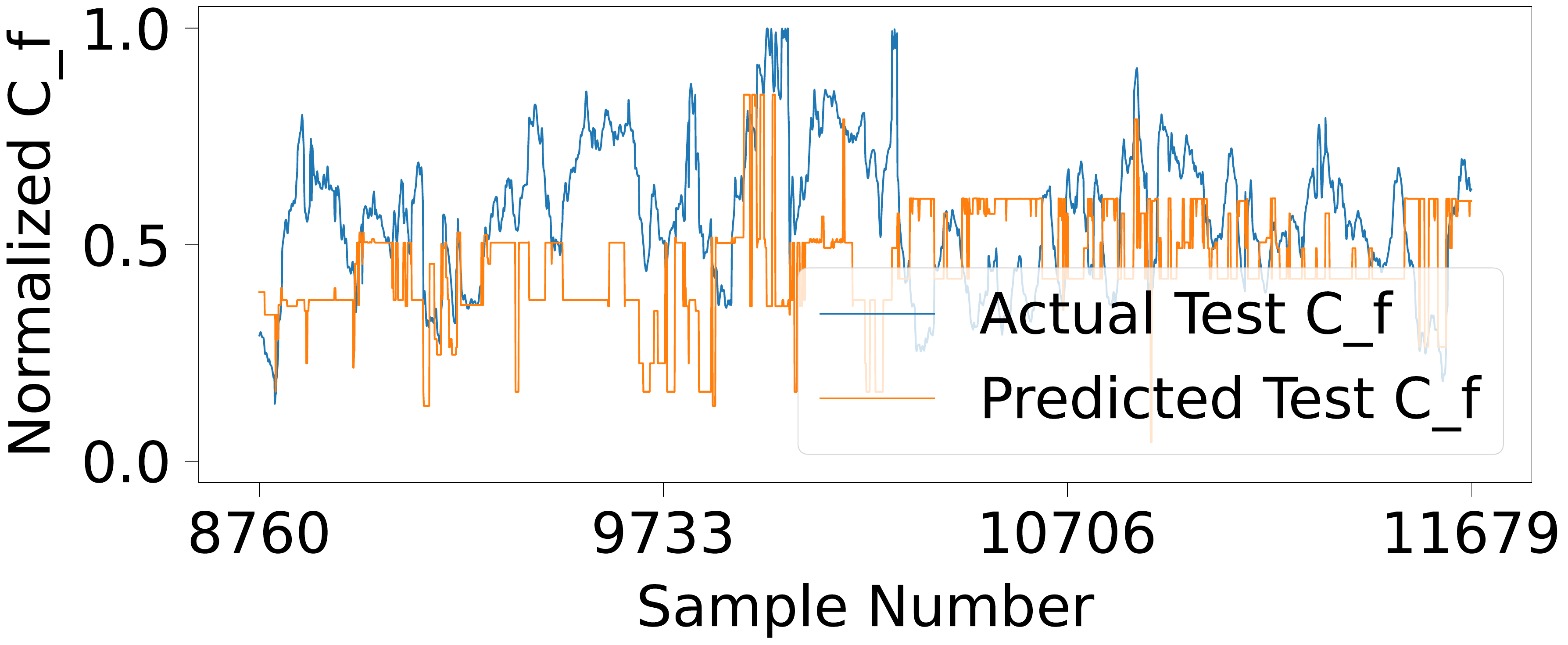}
    \caption{Decision tree}
    \label{fig:dt}
  \end{subfigure}
  \hfill
  \begin{subfigure}[b]{0.495\linewidth}
    \centering
    \includegraphics[width=\linewidth]{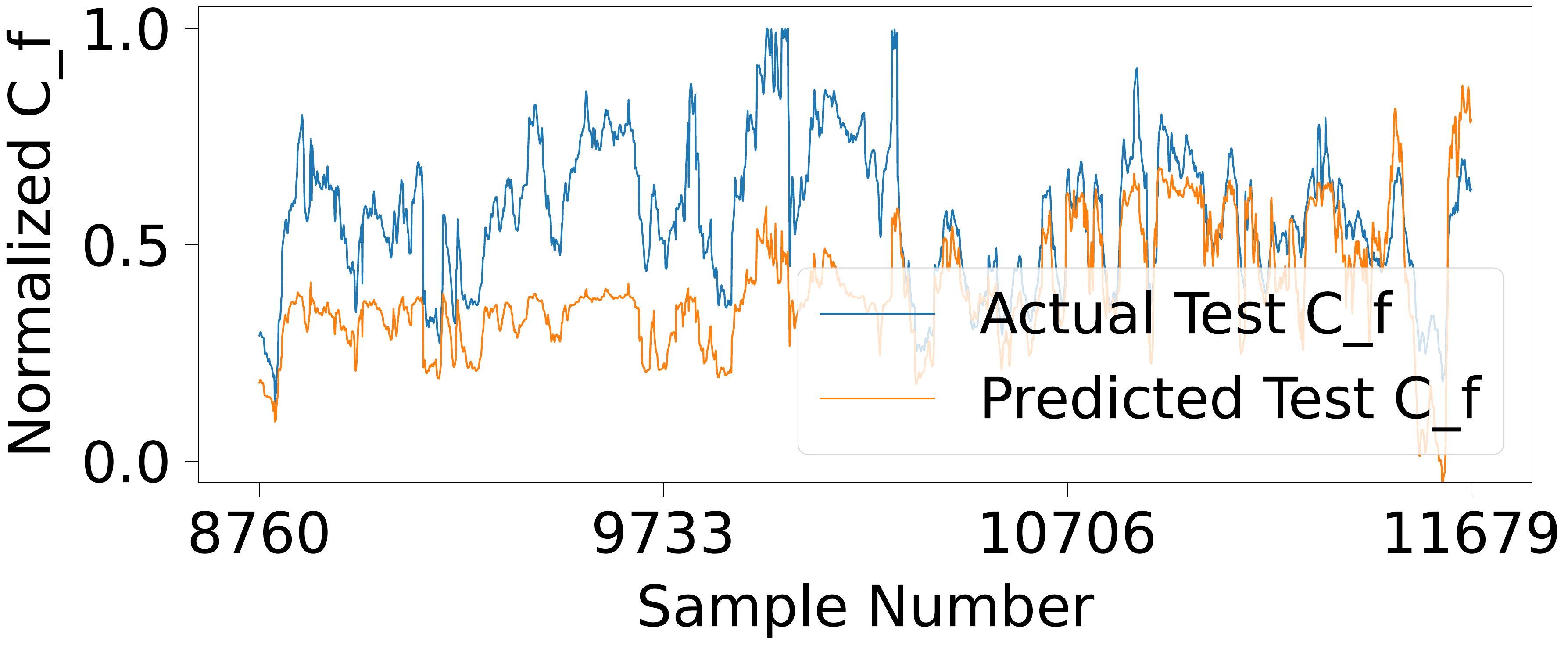}
    \caption{Feed-forward neural network}
    \label{fig:ddnn}
  \end{subfigure}
  
  \vspace{5mm} 
  
  \begin{subfigure}[b]{0.495\linewidth}
    \centering
    \includegraphics[width=\linewidth]{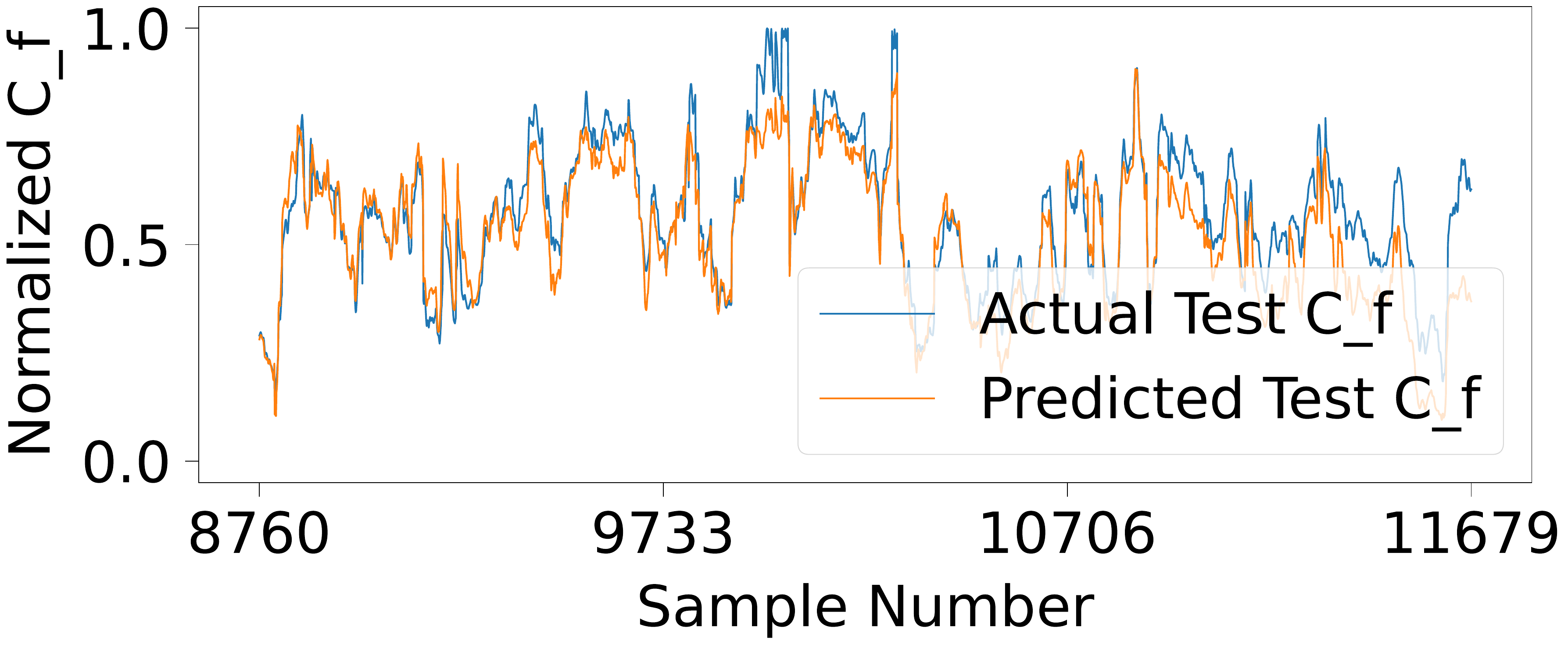}
    \caption{Bidirectional material transfer dynamics PINN}
    \label{fig:pinn1}
  \end{subfigure}
  \hfill
  \begin{subfigure}[b]{0.495\linewidth}
    \centering
    \includegraphics[width=\linewidth]{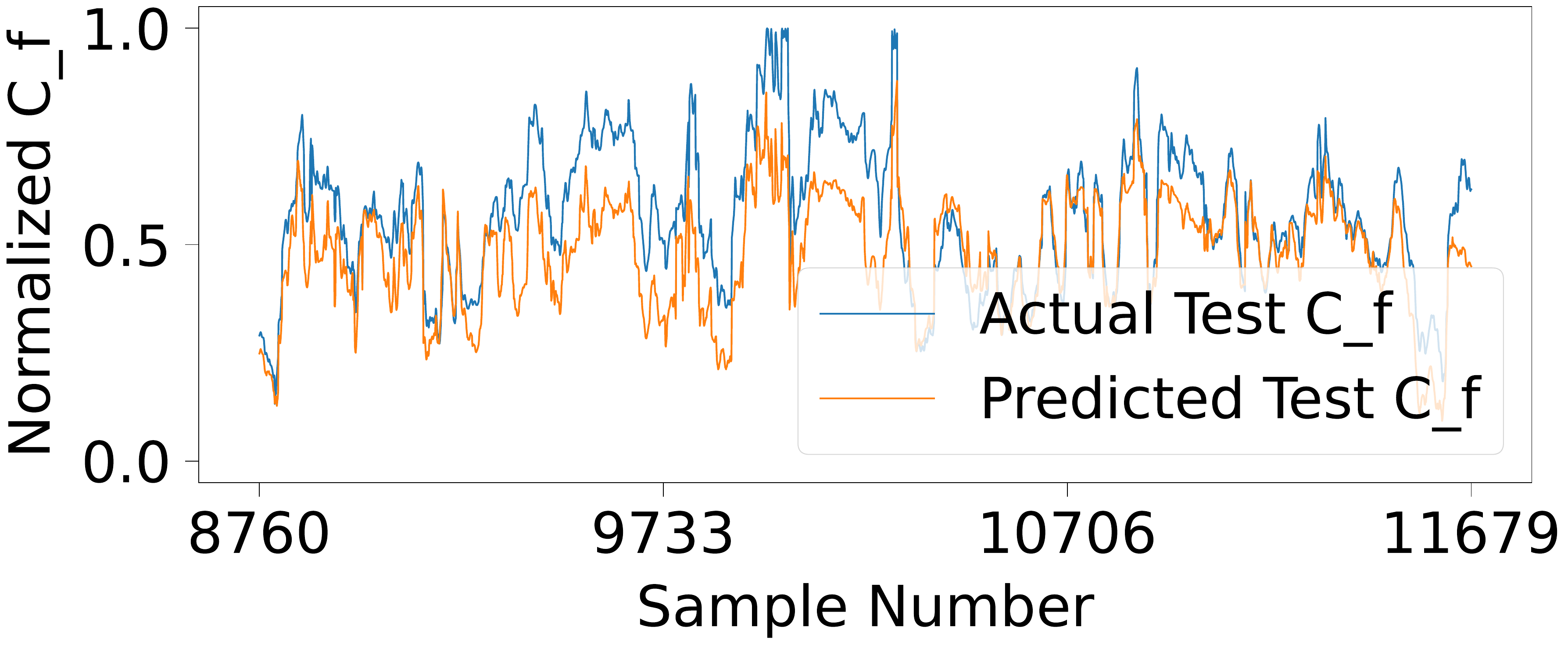}
    \caption{Unidirectional material transfer dynamics PINN}
    \label{fig:pinn2}
  \end{subfigure}

  \vspace{5mm} 

  \begin{subfigure}[b]{0.495\linewidth}
    \centering
    \includegraphics[width=\linewidth]{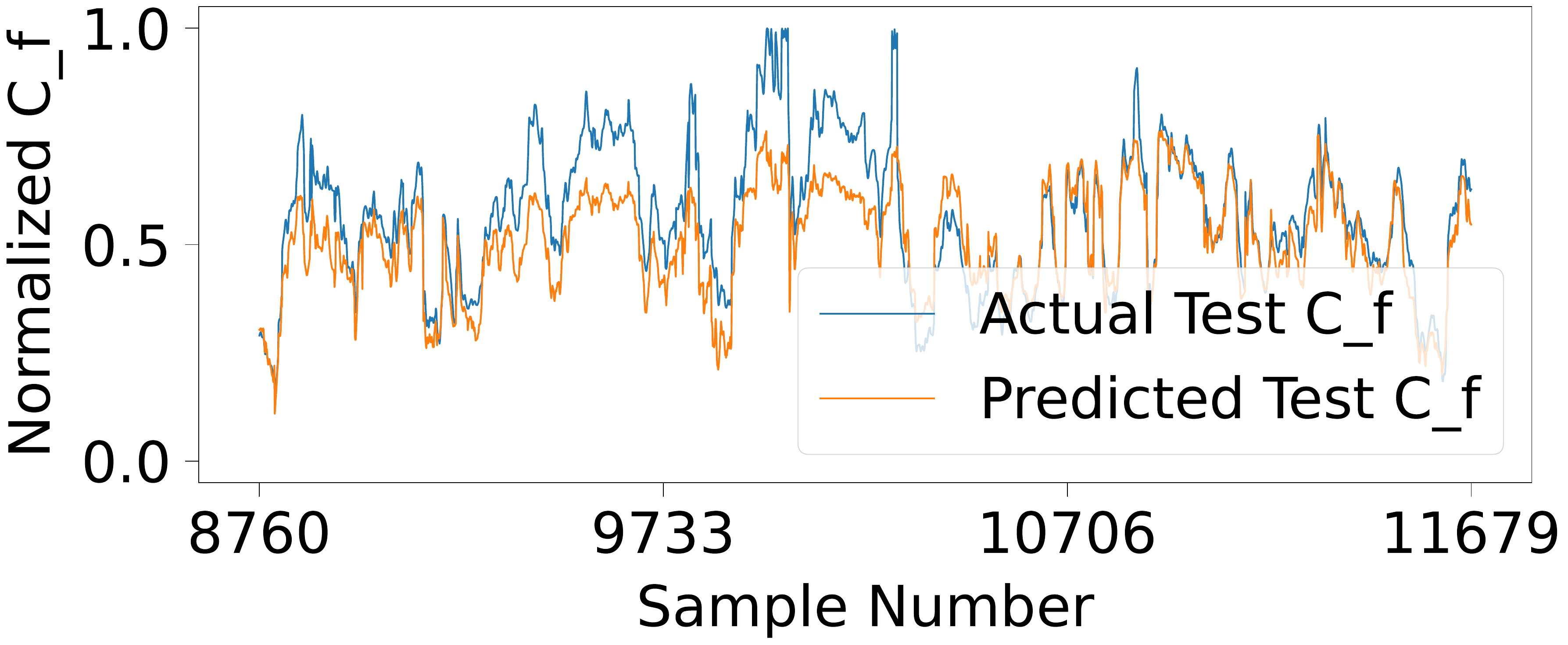}
    \caption{Overall mass balance PINN}
    \label{fig:pinn3}
  \end{subfigure}
  
  \caption{Performance of machine learning models in predicting concentrate gold grade (\( C_f \)) on a subset of the test dataset for Cell I.}
  \label{fig:test_sets_cell3}
\end{figure}

The predictive performance of the physics-informed neural networks, purely data-driven neural network, linear regression, random forest, and decision tree models on a subset of the test dataset for Cell I is depicted in Figure~\ref{fig:test_sets_cell3}. It can be seen from this figure that the predictions from the PINN models closely align with the actual measurements of concentrate gold grade (\( C_f \)) across the test dataset, as showcased in Figure~\ref{fig:test_sets_cell3} (e), (f), and (g), demonstrating their ability to effectively capture the froth flotation dynamics. In contrast, although the purely data-driven models shown in Figure~\ref{fig:test_sets_cell3} (a), (b), (c), and (d) capture some underlying trends, their predictions significantly differ from the actual values. This shows the limitations of the purely data-driven models in accurately modeling the nonlinear and complex dynamics of the froth flotation process.

%
%
%

In Table~\ref{table:MSE_performance_cell3}, the performance of physics-informed neural networks is compared to purely data-driven models for flotation Cell I, in terms of mean squared error (in $(g \cdot t^{-1})^2$) and mean relative error metrics. The results show that PINNs yield better MSE and MRE scores in both validation and testing phases compared to purely data-driven models; however, it should be noted that while the linear regression model achieved a lower validation error than two of the PINN models, its test error is significantly higher than that of all the PINNs. This demonstrates the PINNs' capacity to more accurately model the dynamics of the concentrate gold grade (\(C_f\)), particularly emphasizing their strong generalization ability to new, unseen datasets, as can be noticed from the test MSE/MRE results.

\afterpage{
\begin{table}[h]
\caption{Performance of machine learning models for froth flotation Cell I, evaluated in terms of MSE and MRE metrics on both validation and test datasets.}
\label{table:MSE_performance_cell3}
\renewcommand{\arraystretch}{1.3} 
\begin{tabular}{l@{\hspace{0.28in}}cc@{\hspace{0.28in}}cc}
\hline
& \multicolumn{2}{c@{\hspace{0.38in}}}{\footnotesize Validation} & \multicolumn{2}{c@{\hspace{0.16in}}}{\footnotesize Test} \\
\raisebox{1.8ex}[0pt][0pt]{\footnotesize Machine Learning Model of Cell I} & \footnotesize MSE & \footnotesize MRE & \footnotesize MSE & \footnotesize MRE \\
\hline
\footnotesize Linear regression & \footnotesize 22.032 & \footnotesize 0.071 & \footnotesize 73.741 & \footnotesize 0.144 \\
\footnotesize Random forest & \footnotesize 54.335 & \footnotesize 0.113 & \footnotesize 64.153 & \footnotesize 0.105 \\
\footnotesize Decision tree & \footnotesize 55.864 & \footnotesize 0.109 & \footnotesize 77.960 & \footnotesize 0.122 \\
\footnotesize Feed-forward neural network & \footnotesize 33.402 & \footnotesize 0.083 & \footnotesize 86.641 & \footnotesize 0.135 \\
\footnotesize Bidirectional material transfer dynamics PINN & \footnotesize 25.080 & \footnotesize 0.077 & \footnotesize 35.144 & \footnotesize 0.085 \\
\footnotesize Unidirectional material transfer dynamics PINN & \footnotesize 30.921 & \footnotesize 0.080 & \footnotesize 29.136 & \footnotesize 0.077 \\
\footnotesize Overall mass balance PINN & \footnotesize 17.749 & \footnotesize 0.059 & \footnotesize 22.636 & \footnotesize 0.069 \\
\hline
\end{tabular}
\end{table}
}

\begin{figure}[h]
  \centering
  
  \begin{subfigure}[b]{0.495\linewidth}
    \centering
    \includegraphics[width=\linewidth]{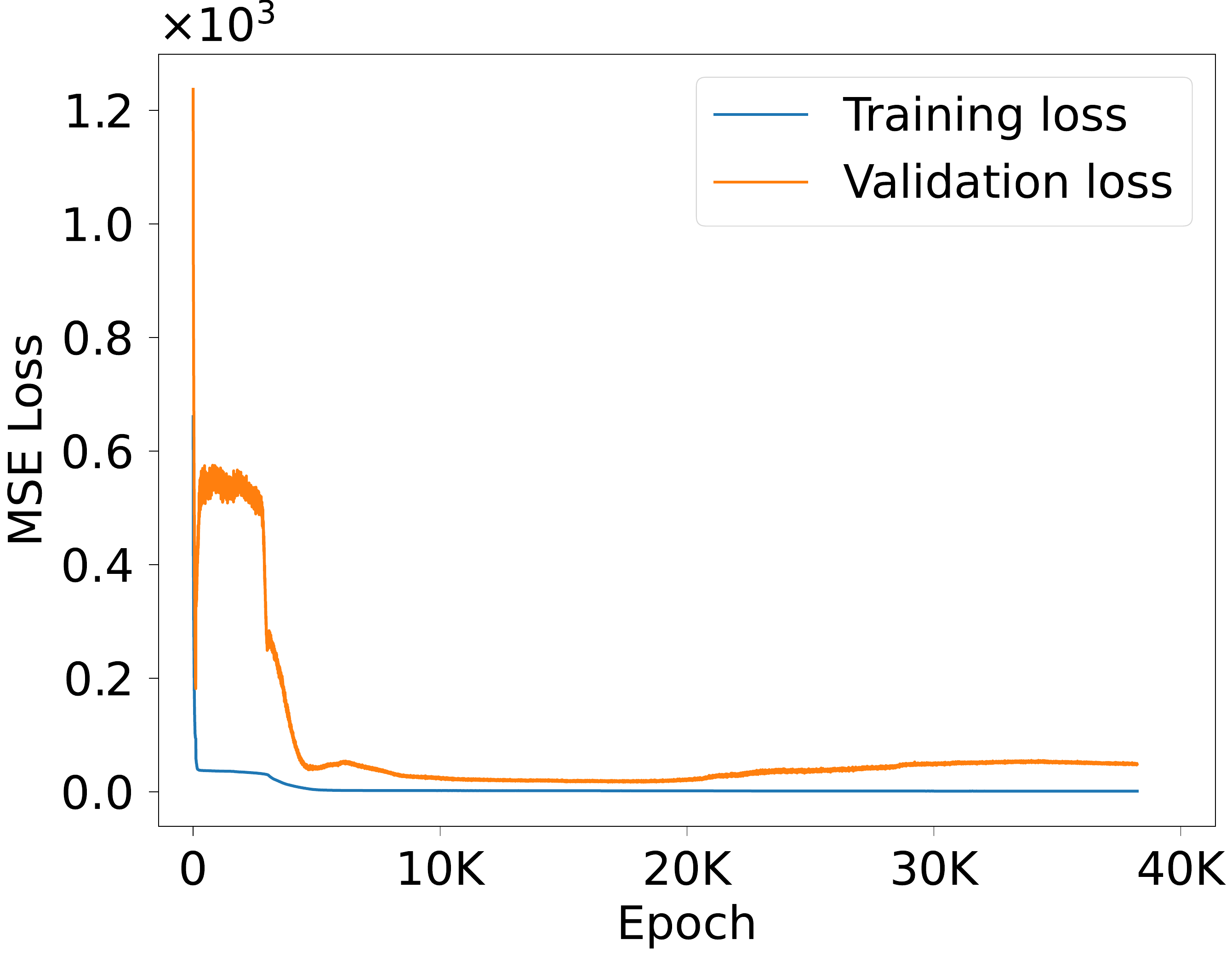}
    \caption{Feed-forward neural network}
    \label{fig:lr}
  \end{subfigure}
  \hfill 
  \begin{subfigure}[b]{0.495\linewidth}
    \centering
    \includegraphics[width=\linewidth]{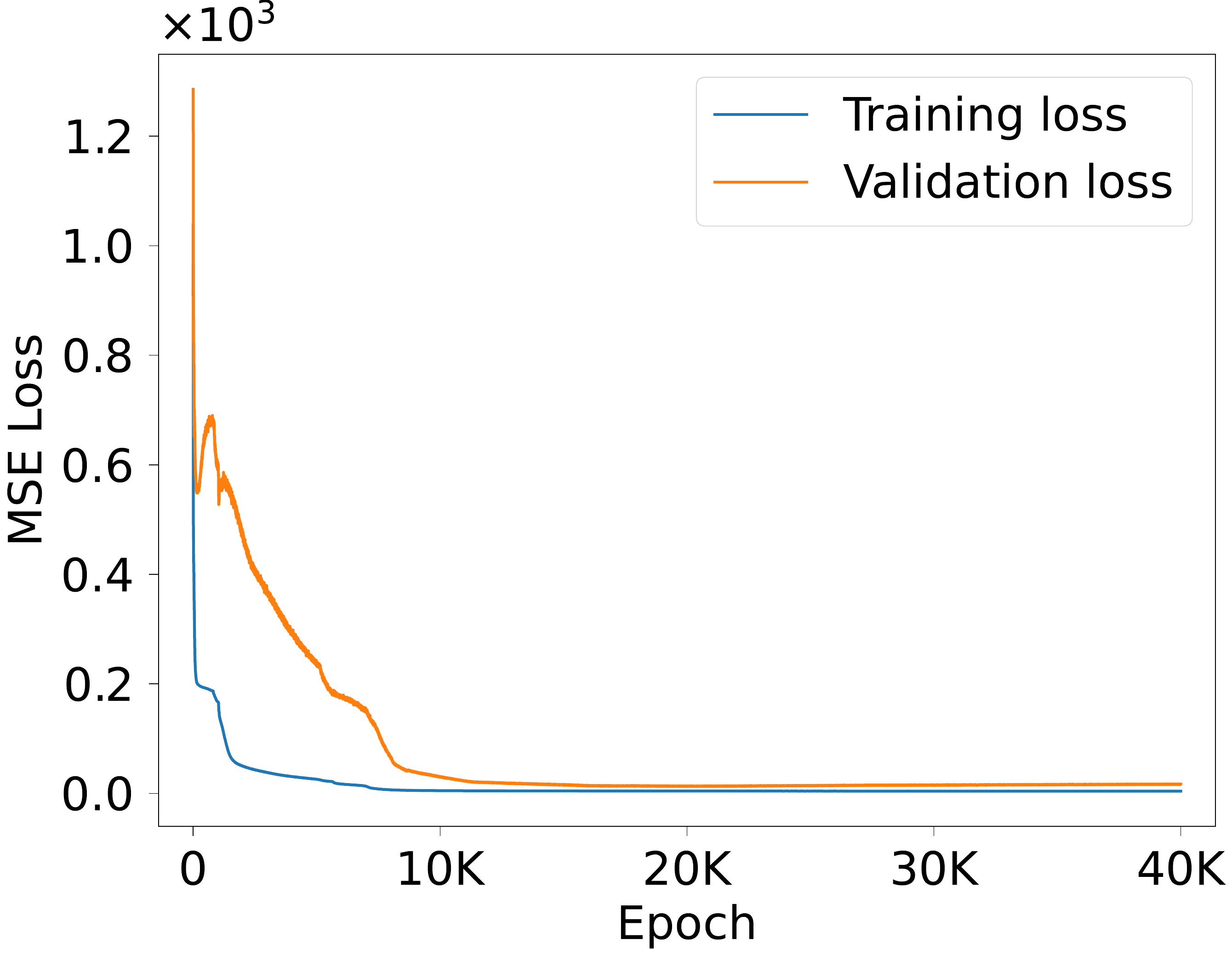}
    \caption{Bidirectional material transfer dynamics PINN}
    \label{fig:rf}
  \end{subfigure}
  
  \vspace{5mm} 
  
  \begin{subfigure}[b]{0.495\linewidth}
    \centering
    \includegraphics[width=\linewidth]{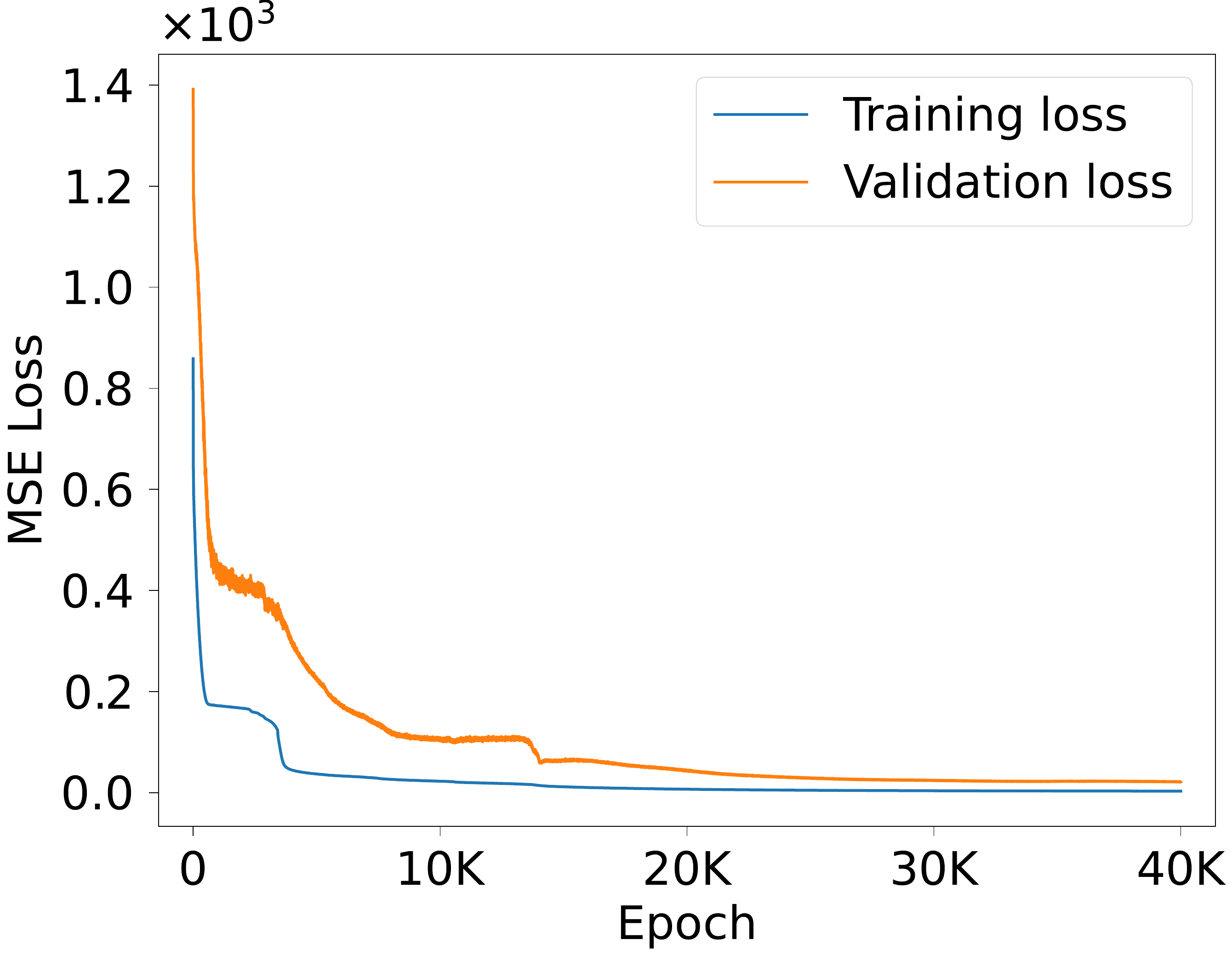}
    \caption{Unidirectional material transfer dynamics PINN}
    \label{fig:dt}
  \end{subfigure}
  \hfill
  \begin{subfigure}[b]{0.495\linewidth}
    \centering
    \includegraphics[width=\linewidth]{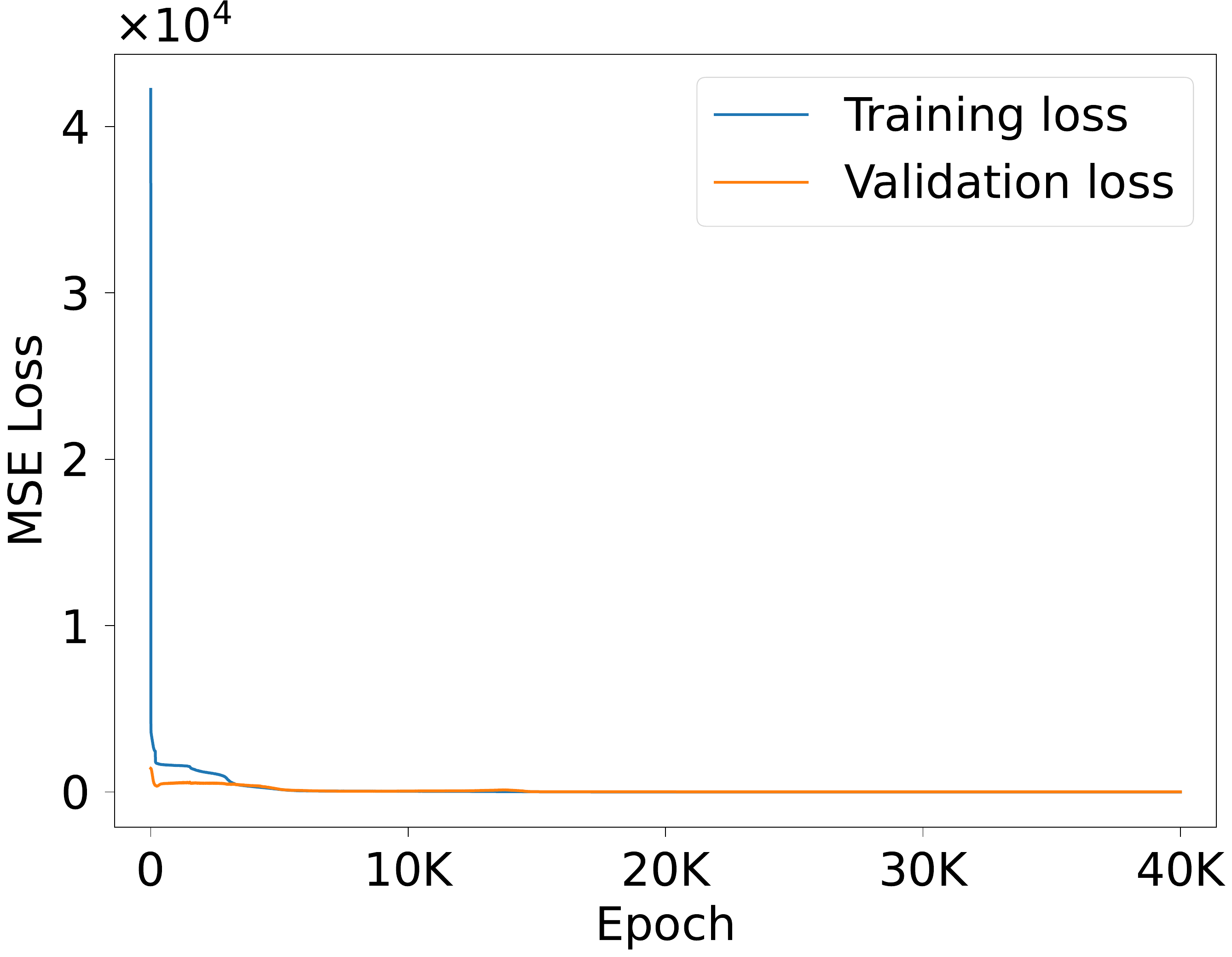}
    \caption{Overall mass balance PINN}
    \label{fig:ddnn}
  \end{subfigure}
  
  \caption{Loss evolution of training and validating of neural network models for flotation Cell I.}
  \label{fig:loss_cell3}
\end{figure}

The evolution of loss for both the physics-informed and purely data-driven neural networks employed to model the dynamics of flotation Cell I is presented in Figure~\ref{fig:loss_cell3}. This figure illustrates the convergence and learning progress during the training and validation phases. Specifically, Figure~\ref{fig:loss_cell3} (a) shows the initial rapid decrease in loss for the purely data-driven neural network during the early training epochs. However, as training progresses, the validation loss increases and the performance of the model on the validation set deteriorates, indicative of an overfitting issue. In contrast, Figure~\ref{fig:loss_cell3} (b), (c), and (d) depict the loss trends for the PINN models, which achieve a steep decline in both training and validation losses during the early training epochs and demonstrate stability after a certain number of epochs. Unlike the purely data-driven neural network, the patterns of validation loss in the PINN models demonstrate their strong generalization capabilities beyond the training data, illustrating their potential to mitigate overfitting to the training set.

%
%
%

From the analysis above it can be concluded that physics-informed neural networks outperform purely data-driven models in terms of predictive accuracy and generalization to unseen data in froth flotation Cell I. By incorporating even partially known physical laws into their learning processes, PINNs are capable of delivering more accurate predictions in varying industrial environments, where data may be sparse and noisy. Leveraging both empirical data and physical principles, these models not only generate more accurate predictions but also ensure that the solutions are physically plausible, thereby providing insights into the optimization process.

%
\subsection{Predictive Analysis for Flotation Cell II}

Following the methodology applied in Cell I, the analysis for Cell II similarly leveraged three distinct datasets, each sampled during different time periods for the training, validation, and testing of the models. Specifically, the training dataset for Cell II included 17551 data points, the validation dataset comprised 9157 points, and the test dataset contained 12430 points, each including the 14 variables as listed in Table~\ref{table:flotation_cell_variables}. As with Cell I, these data points are distributed across the spatio-temporal domain.

The feed-forward neural network \( \mathbf{u}(t, \mathbf{x}) \) employed for Cell II includes three hidden layers with 128, 256, and 128 neurons, respectively. The configuration and hyperparameters, such as loss and activation functions, optimizer, learning rate, and volume constraints, are identical to those used in the models for Cell I, unless otherwise specified. Additionally, the auxiliary neural network \( R(t,\mathbf{x}, \mathbf{u}(t,\mathbf{x})) \), which is designed to estimate the average flotation rate \( R \) for the unidirectional material transfer dynamics PINN model, consists of a single hidden layer with 400 neurons. For Cell II, the hyperparameters of all the machine learning models were also calibrated during the training phase to ensure accurate predictions of the output variable of interest \( C_p \). The hyperparameters, the number of hidden layers and their respective number of neurons for the neural networks modeling Cell II are detailed in Table \ref{table:hyperparameter_cell4}. Moreover, early stopping was implemented for all neural network models, configured with a patience of 30000 and a tolerance parameter of \( 10^{-5} \).

\begin{table}[h]
\caption{Neural network configurations and hyperparameters for modeling flotation Cell II.}
\label{table:hyperparameter_cell4}
\renewcommand{\arraystretch}{1.3} 
\begin{tabular}{l@{\hspace{0.9in}}l}
\hline
\textbf{Hyperparameter} & \textbf{Configuration} \\
\hline
Optimizer & Adam \\
Activation function & Tanh \\
Learning rate & $10^{-5}$ \\
Batch size & $128$ \\
\hline
\multicolumn{2}{l}{\textbf{Network \( \mathbf{u}(t, \mathbf{x}) \) Architecture}} \\
\hline
Number of hidden layers & 3 \\
Number of neurons in the hidden layers & 128, 256, 128 \\
\hline
\multicolumn{2}{l}{\textbf{Network \( R(t,\mathbf{x}, \mathbf{u}(t,\mathbf{x})) \) Architecture}} \\
\hline
Number of hidden layer & 1 \\
Number of neurons in the hidden layer & 400 \\
\hline
\end{tabular}
\end{table}

\begin{figure}[htbp]
  \centering
  
  \begin{subfigure}[b]{0.495\linewidth}
    \centering
    \includegraphics[width=\linewidth]{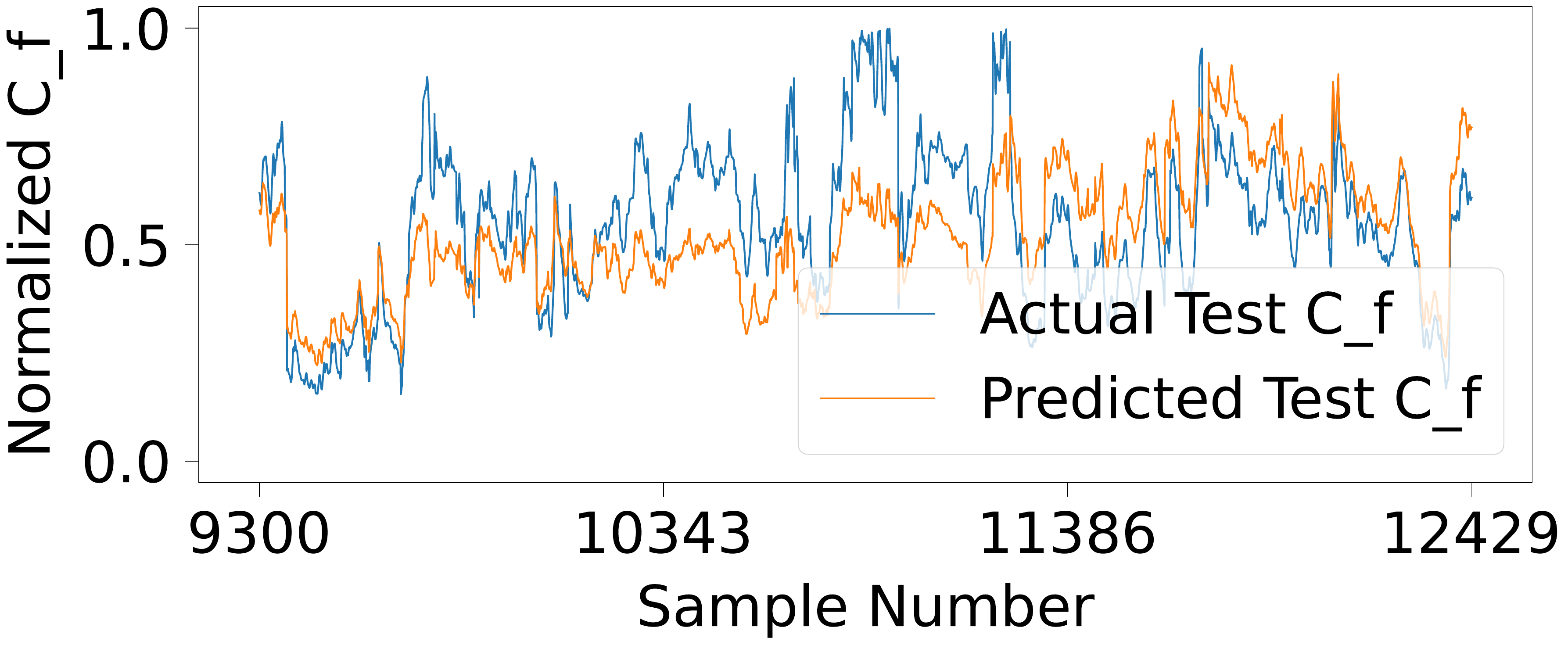}
    \caption{Linear regression}
    \label{fig:lr}
  \end{subfigure}
  \hfill 
  \begin{subfigure}[b]{0.495\linewidth}
    \centering
    \includegraphics[width=\linewidth]{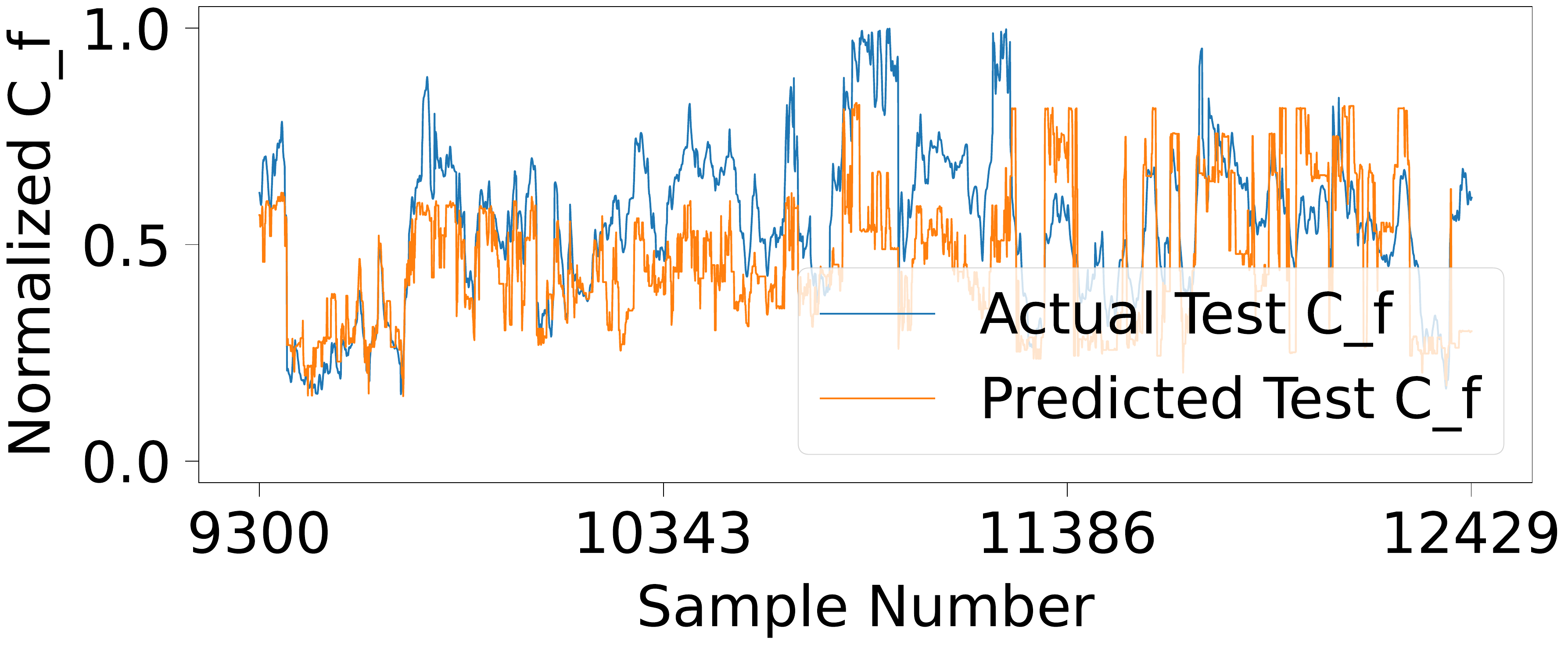}
    \caption{Random forest}
    \label{fig:rf}
  \end{subfigure}
  
  \vspace{5mm} 
  
  \begin{subfigure}[b]{0.495\linewidth}
    \centering
    \includegraphics[width=\linewidth]{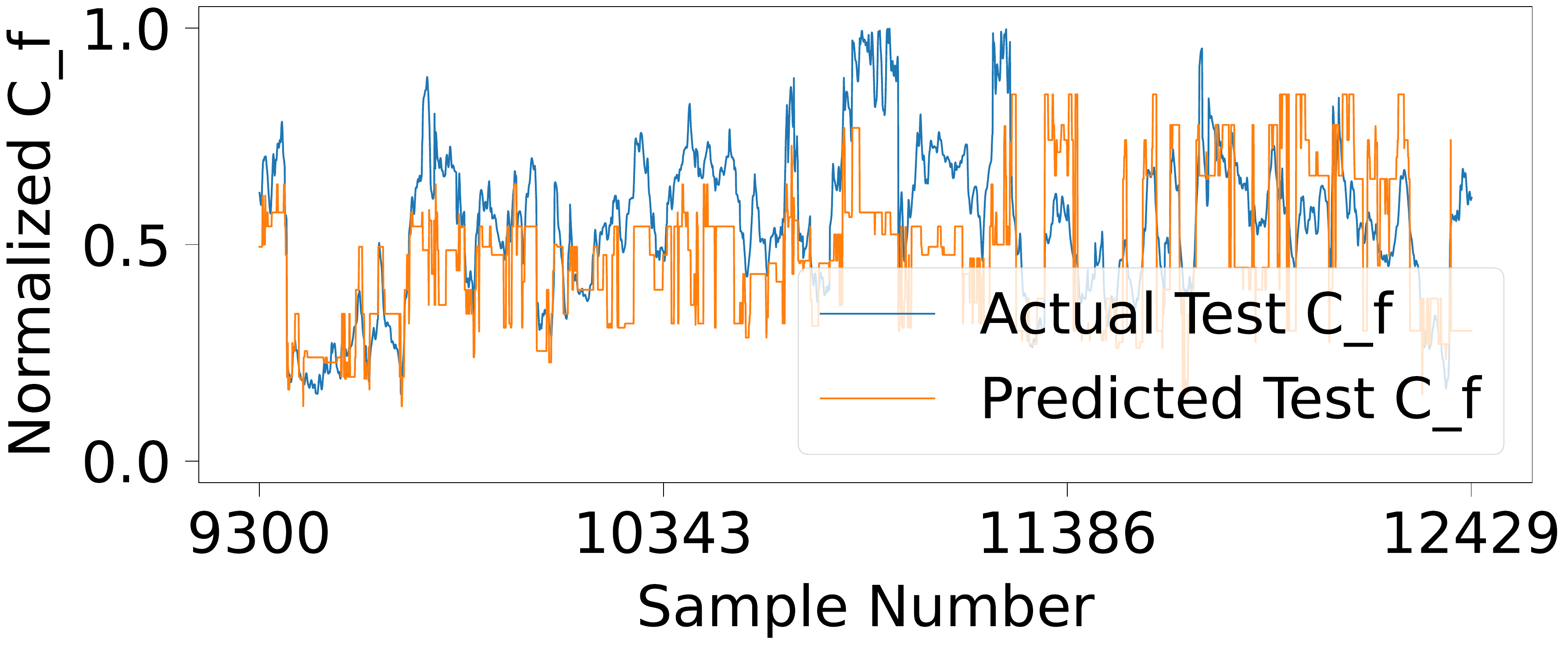}
    \caption{Decision tree}
    \label{fig:dt}
  \end{subfigure}
  \hfill
  \begin{subfigure}[b]{0.495\linewidth}
    \centering
    \includegraphics[width=\linewidth]{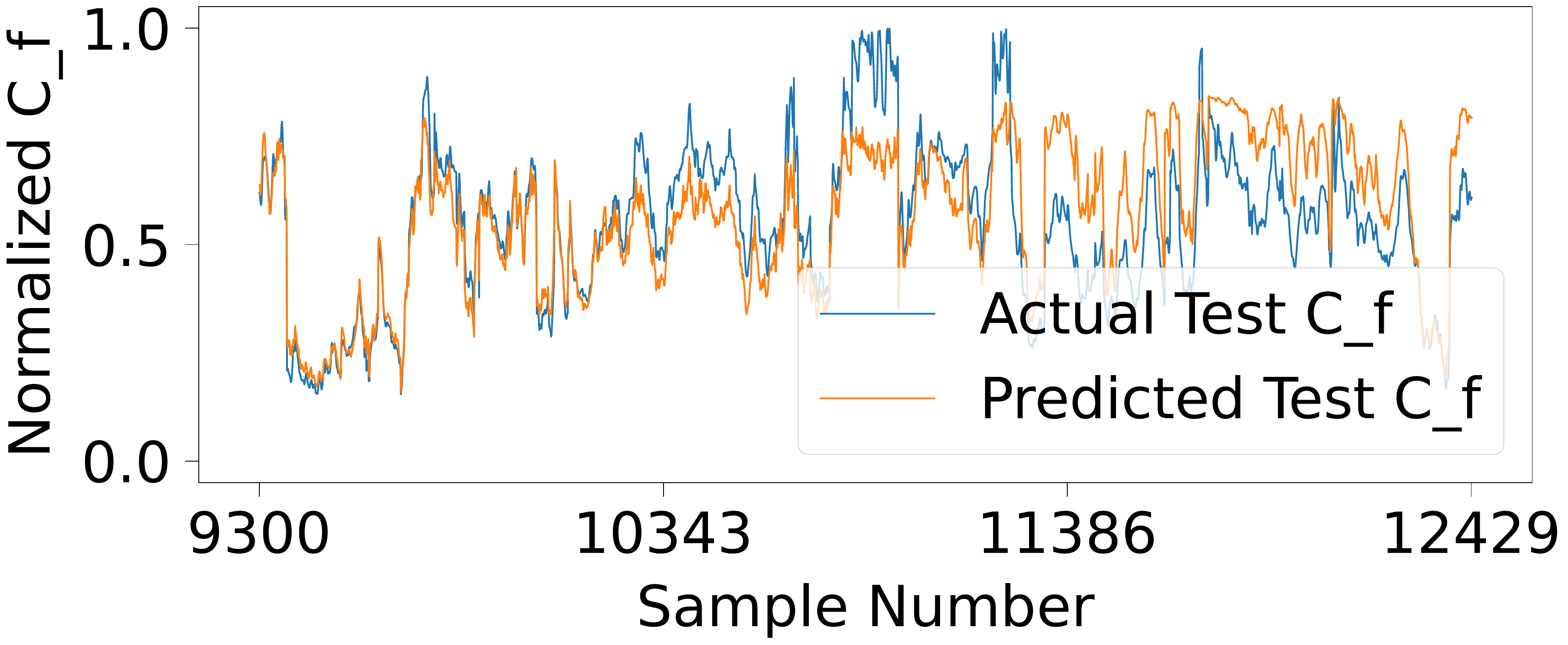}
    \caption{Feed-forward neural network}
    \label{fig:ddnn}
  \end{subfigure}
  
  \vspace{5mm} 
  
  \begin{subfigure}[b]{0.495\linewidth}
    \centering
    \includegraphics[width=\linewidth]{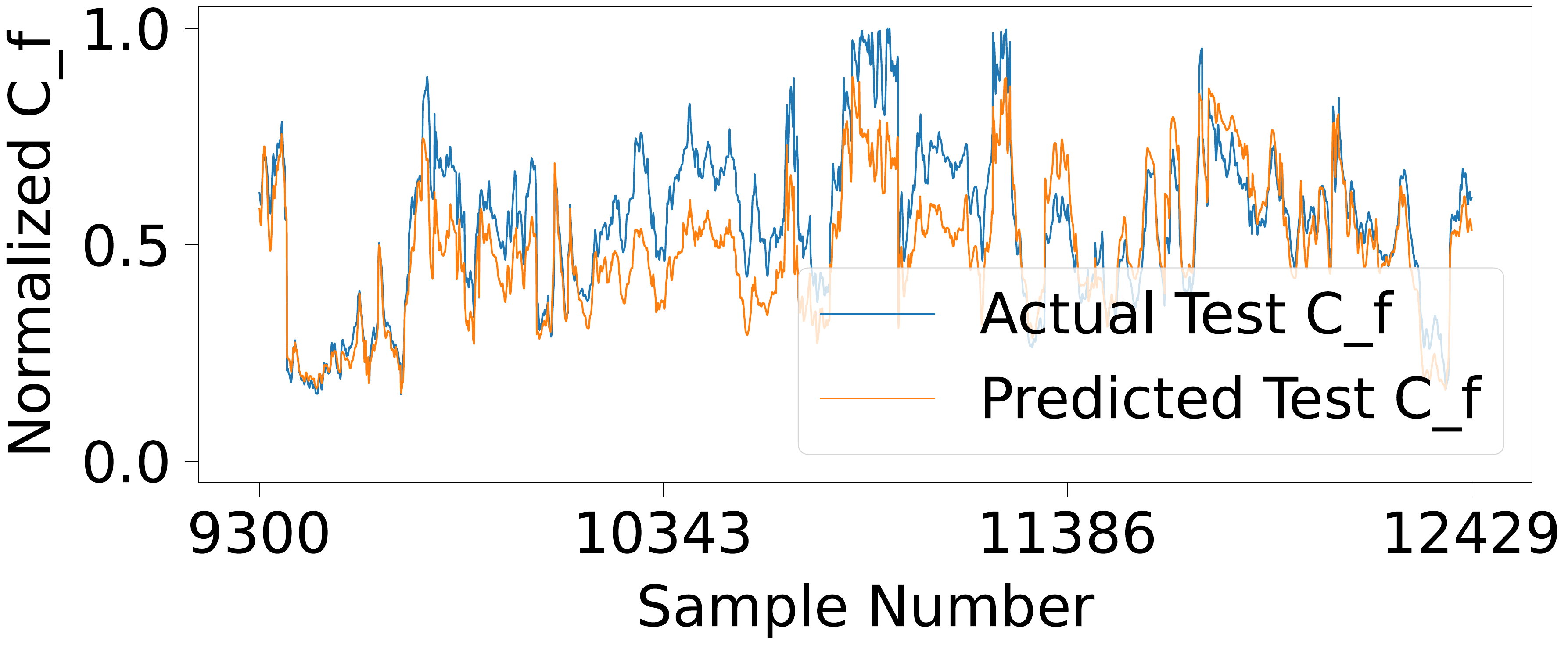}
    \caption{Bidirectional material transfer dynamics PINN}
    \label{fig:pinn1}
  \end{subfigure}
  \hfill
  \begin{subfigure}[b]{0.495\linewidth}
    \centering
    \includegraphics[width=\linewidth]{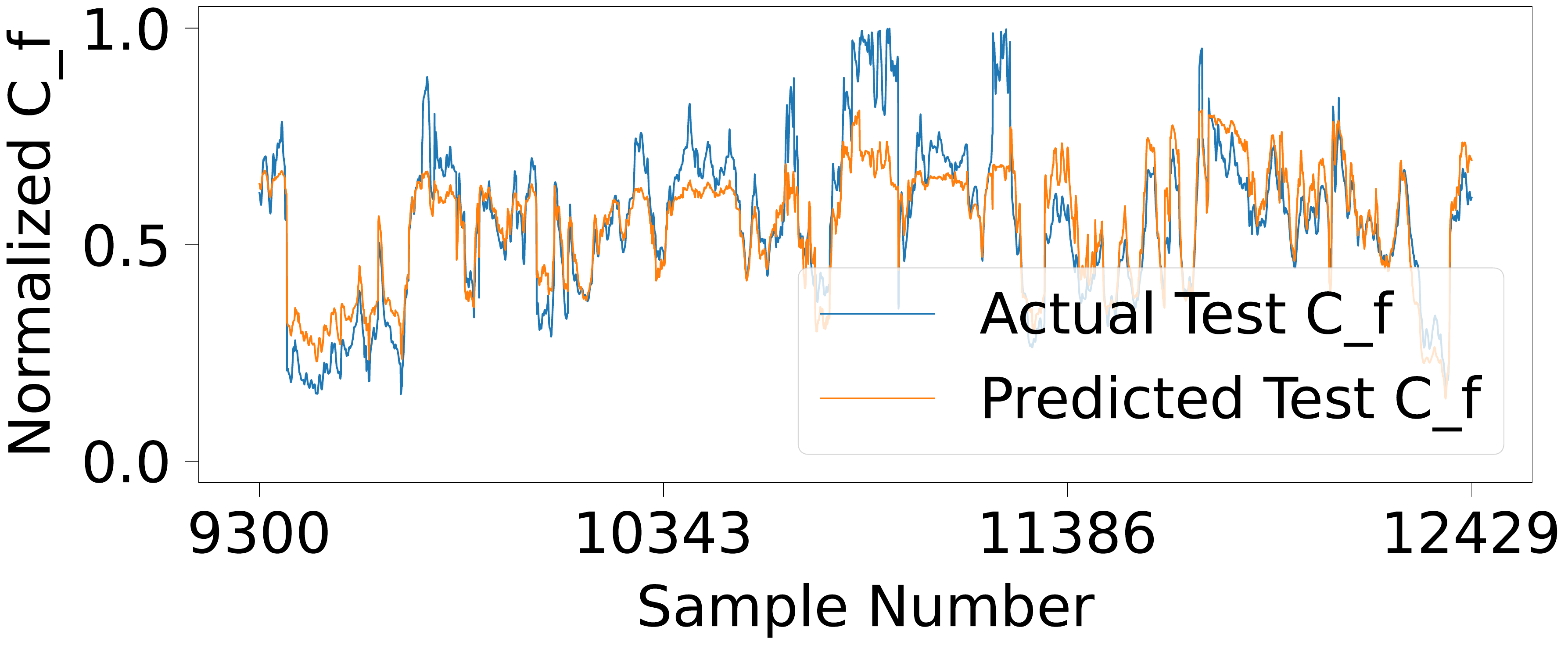}
    \caption{Unidirectional material transfer dynamics PINN}
    \label{fig:pinn2}
  \end{subfigure}

  \vspace{5mm} 

  \begin{subfigure}[b]{0.495\linewidth}
    \centering
    \includegraphics[width=\linewidth]{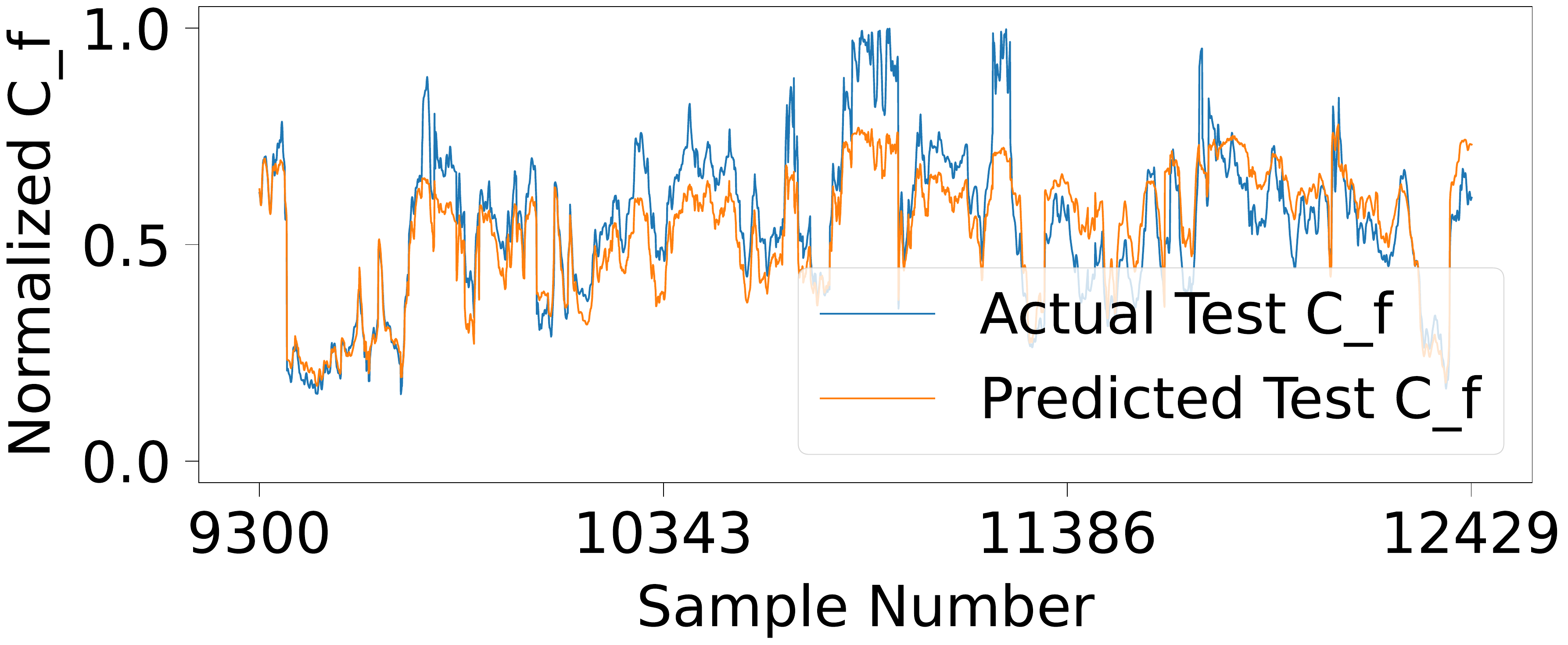}
    \caption{Overall mass balance PINN}
    \label{fig:pinn3}
  \end{subfigure}
  
  \caption{Performance of machine learning models in predicting concentrate gold grade (\( C_f \)) on a subset of the test dataset for Cell II.}
  \label{fig:test_sets_cell4}
\end{figure}

Figure~\ref{fig:test_sets_cell4} illustrates the predictive performance of the physics-informed neural networks, purely data-driven neural network, linear regression, random forest, and decision tree for flotation Cell II across a subset of the test dataset. The PINN models are capable of accurately approximating the actual measurements of the concentrate gold grade (\(C_f\)), and the pattern of the predictions closely matches the actual pattern, as shown in Figure~\ref{fig:test_sets_cell4} (e), (f), and (g). In contrast, the purely data-driven models, depicted in Figure~\ref{fig:test_sets_cell4} (a), (b), (c), and (d), only partially capture the pattern, with their predictions often deviating significantly from the actual concentrate gold grade. This deviation shows the limited capability of data-driven models and the challenges they face in modeling the nonlinear and complex dynamics of the froth flotation process.

%
%
%

The performance of physics-informed neural networks and purely data-driven machine learning models for flotation Cell II is also quantified using both the mean squared error (in $(g \cdot t^{-1})^2$) and mean relative error metrics, as depicted in Table~\ref{table:MSE_performance_cell4}. The evaluation demonstrates that PINNs significantly outperform purely data-driven models in both validation and test MSE and MRE scores. Such performance showcases the ability of PINNs to more accurately model the dynamics of the concentrate gold grade (\(C_f\)). Specifically, the lower error scores on the test set further illustrate the PINNs' robust generalization to unseen data.

\afterpage{
\begin{table}[htbp]
\caption{Performance of machine learning models for froth flotation Cell II, evaluated in terms of MSE and MRE metrics on both validation and test datasets.}
\label{table:MSE_performance_cell4}
\renewcommand{\arraystretch}{1.3} 
\begin{tabular}{l@{\hspace{0.28in}}cc@{\hspace{0.28in}}cc}
\hline
& \multicolumn{2}{c@{\hspace{0.38in}}}{\footnotesize Validation} & \multicolumn{2}{c@{\hspace{0.16in}}}{\footnotesize Test} \\
\raisebox{1.8ex}[0pt][0pt]{\footnotesize Machine Learning Model of Cell II} & \footnotesize MSE & \footnotesize MRE & \footnotesize MSE & \footnotesize MRE \\
\hline
\footnotesize Linear regression & \footnotesize 13.260 & \footnotesize 0.090 & \footnotesize 27.893 & \footnotesize 0.131 \\
\footnotesize Random forest & \footnotesize 12.424 & \footnotesize 0.078 & \footnotesize 22.479 & \footnotesize 0.094 \\
\footnotesize Decision tree & \footnotesize 16.949 & \footnotesize 0.095 & \footnotesize 25.552 & \footnotesize 0.102 \\
\footnotesize Feed-forward neural network & \footnotesize15.542 & \footnotesize 0.089 & \footnotesize 23.489 & \footnotesize 0.092 \\
\footnotesize Bidirectional material transfer dynamics PINN & \footnotesize 8.190 & \footnotesize 0.059 & \footnotesize 16.496 & \footnotesize 0.081 \\
\footnotesize Unidirectional material transfer dynamics PINN & \footnotesize 8.741 & \footnotesize 0.063 & \footnotesize 13.886 & \footnotesize 0.082 \\
\footnotesize Overall mass balance PINN & \footnotesize 10.563 & \footnotesize 0.073 & \footnotesize 11.758 & \footnotesize 0.065 \\
\hline
\end{tabular}
\end{table}
}

It should be emphasized that the success of the neural networks optimization is significantly influenced by their architectures and the selection of the learning rate. Although it is important for the architecture to be sufficiently expressive to capture the system's dynamics, an overly complex architecture should be avoided to prevent excessive computational burdens during the training procedure. On the other hand, for the neural network models discussed in this study, a learning rate of $10^{-5}$ was set based on empirical observations during the training process. A higher learning rate was found to lead to overshooting issues, while a lower rate caused the models to become stuck in local minima.

\section{Conclusions}
\label{app:concl}


In this paper, physics-informed neural network models were developed to predict the concentrate gold grade in two flotation cells. Specifically, three mathematical models, expressed as ordinary differential equations, along with domain knowledge, were integrated into conventional deep learning methods to model the complex dynamics of flotation processes. The results showed that in test scenarios, physics-informed models outperformed purely data-driven machine learning models in mean squared error and mean relative error metrics, indicating the robust generalization capabilities of the PINNs to unseen data. Additionally, the evolution of training and validation losses illustrates that PINNs effectively prevent overfitting to the training data, unlike the purely data-driven models, which exhibited increased validation loss as training progressed, indicative of capturing noise and therefore overfitting. In conclusion, the integration of even partially known physical laws into the learning algorithms enables physics-informed neural networks to not only make accurate predictions under varying conditions and in environments where data may be sparse and noisy but also produce physically plausible solutions, thereby enhancing the understanding of the optimization problem.

It should be noted that the most fundamental mathematical model employed in overall mass balance PINN, has been infrequently explored within froth flotation research. While this classical model has been used to develop various kinetic models, its direct application in capturing the dynamics of the froth flotation process has been limited. However, the physics-informed neural network that incorporates this mathematical model, demonstrates superior performance compared to the other PINN models, despite its simplicity.

The research presented here illustrates the efficacy of physics-informed machine learning as a viable alternative to traditional, purely data-driven or solely physics-based models in the froth flotation process. Additionally, it establishes a foundational framework that could guide future research efforts aimed at enhancing the performance of industrial flotation cells.

\section*{Credit Authorship Contribution Statement}

\textbf{Mahdi Nasiri}: Conceptualization, Writing – original draft, Software, Writing – review \& editing. \textbf{Sahel Iqbal}: Conceptualization, Supervision, Writing – review \& editing. \textbf{Simo Särkkä}: Conceptualization, Supervision, Writing – review \& editing.

\section*{Declaration of Competing Interest}

The authors declare that they have no known competing financial interests or personal relationships that could have appeared to influence the work reported in this paper.

\section*{Data Availability}

The data used in this study are confidential.

\section*{Acknowledgements}

The authors acknowledge the funding provided by Business Finland through the project "Development of Artificial Intelligence and Machine Learning for Online Perception and Operating Mode Optimization in Process Industry". The authors also extend their appreciation to Cesar Araujo for his detailed comments on flotation dynamics modeling.

\bibliographystyle{elsarticle-harv} 
\bibliography{references}
\end{document}